\documentclass[twocolumn,prb]{revtex4-2}
\usepackage[usenames,dvipsnames]{color}
\usepackage{comment}
\usepackage{graphicx}
\usepackage[utf8x,utf8]{inputenc}
\usepackage{lineno}
\usepackage{color}
\usepackage{hyperref}
\usepackage{array}
\usepackage{amsmath}
\usepackage{braket}
\usepackage{caption}
\usepackage{subcaption}
\usepackage{setspace}
\usepackage[section]{placeins} 
\usepackage{multirow}
\usepackage{booktabs}
\usepackage{xcolor}
\usepackage{amssymb}
\usepackage{cancel}

\let\oldchi\chi

\renewcommand{\chi}{\scalebox{1.1}{$\oldchi$}}

\usepackage[]{todonotes}

\begin{document}

\title{ Pairing susceptibility in the weakly interacting multilayer Hubbard model evaluated by direct perturbative expansion  }

\author{Rayan Farid}
\email{rfarid@mun.ca}
\affiliation{Department of Physics and Physical Oceanography, Memorial University of Newfoundland, St. John's, Newfoundland \& Labrador, Canada A1B 3X7}
\author{J. P. F. LeBlanc}
\email{jleblanc@mun.ca}
\affiliation{Department of Physics and Physical Oceanography, Memorial University of Newfoundland, St. John's, Newfoundland \& Labrador, Canada A1B 3X7} 

\date{\today}
\begin{abstract}
We present a systematic study of the interaction, doping, and layer dependence of the $d_{x^2-y^2}$-wave pairing susceptibility of the Hubbard model for a stacked 2D square lattice. We perform a multi-index perturbative expansion up to fourth-order to obtain coefficients in powers of the Hubbard $U$, the inter-layer $V$, and the pair-hopping $J$ interactions. We evaluate the vertex diagrams that contribute to the pairing susceptibility for $\ell= 2,3, 4$ layered models in the $U$-$V$-$J$ interaction space. This provides unprecedented access to the pairing amplitudes, allowing us to identify the processes that enhance or reduce pairing. We distinguish pairing within the diagonal channel, $P^{\parallel}_{d}$, and off-diagonal channel, $P^{\perp}_{d}$, and find that, in the absence of $J$, the qualitative behavior of the layered system is equivalent to the single-layer model. In the presence of $J$, we show that pairing is enhanced sublinearly with increasing $\ell$ and is primarily mediated by the $P^{\perp}_{d}$ component and find which coefficients and diagram sets are responsible. Finally, we construct a generalized $\ell$-dependent equation for $ P^{\perp}_{d}$ to speculate pairing beyond $\ell=4$.

\end{abstract}
\maketitle

\section{Introduction}
Understanding the mechanisms that give rise to superconductivity for models or materials has been a lofty goal within condensed matter theory and the focus of numerous studies. In the case of strongly correlated models, this pursuit has driven advancement in numerical methods,\cite{benchmarks,hchains,no-supercond,stripes,materials,hubbardreview,Schaefer:2020} with the Hubbard model playing a special role. This is due to its relevant early application to understanding cuprates\cite{coldea:2001,leblanc:2019} (LSCO) and the similarities between the Hubbard phase diagram and the cuprate phase diagram\cite{scphase,resistivityphase} both for finite systems\cite{dcamott,gull:2013,haule:2011} and in the ground state\cite{boxiao:2016, Deng14}.
Despite successes, there remains a large gap between the physics of the 2D Hubbard model and the complexity of layered cuprate materials. Further, the local assumption of the Hubbard interaction seems to be insufficient to describe superconducting ground states, and it has been shown that non-local attraction gives rise to pairing and delicate interplay between charge-density wave and superconducting ground states, stripes, and other exotic $T=0$ phases.\cite{jiang:groundstate,jiang:4leg,tang:2024}  
At finite temperatures, the importance of non-local interactions is also seen, and together, these ideas motivate us to expand our considerations beyond local interactions.\cite{jiang:scnnV,Gazizova(2023), peng:2023,3band_Pd}  
For Hubbard-like models, adding to the interaction space increases the computational complexity exponentially, akin to a multi-band problem. Interestingly, there is no fundamental difference between a general multi-band problem and one arising by coupling two or more 2D planes.

In this work, we study $d$-wave pairing processes on infinite 2D planes in the paramagnetic state coupled via kinetic hopping, onsite and interplane repulsion, as well as both local and non-local pair hopping processes by computing the uniform correlated pairing susceptibility $P_d(\mathbf{Q}=0)$ that is nothing more than the vertex contributions of the zero-frequency particle-particle susceptibility projected into an appropriate symmetry channel. We do this via a perturbative approach using the state-of-the-art symbolic tool Algorithmic Matsubara Integration (AMI).\cite{AMI,GIT} Initially, we produce the symbolic representation of all the particle-particle topologies.\cite{farid:2023} Then, we iteratively assign the band index to each diagram symbolically, generating a complete set of particle-particle diagram sets and the Matsubara sums are then resolved by repeated application of the residue theorem. The spatial dependency is stochastically integrated out, ensuring that our results are within the thermodynamic limit and do not suffer from finite-size effects. We perturbatively evaluate $P_d$ in an order-by-order basis up to the fourth order and restrict to a high temperature $\beta t=5$ and weak coupling regime $U/t\leq 3$ to ensure high order corrections are minimum. 

We categorize the computed diagram sets based on the type and number of interaction lines present, resulting in coefficients for powers of $U,V,$ and $J$. These coefficients provide full access to the interaction space without additional computational expense, an advantage unique to our symbolic approach. Leveraging this, we extensively probe the interaction space and perform a comparative analysis of the dominant coefficients to identify diagrammatic processes that enhance or suppress pairing. We find that the presence of a local interaction has no qualitative impact on coupled planes compared to a single 2D plane. With the inclusion of local and non-local pair hopping processes, layer-dependent features begin to appear within our model, where the $P_{d}$ changes sublinearly as a function of layer number. In an instructive manner, we extensively explore this up to a four-layered system as a function of fixed chemical potential $\mu$ and identify the diagram contributions to each pairing channel. In the supplemental materials, we examine further the role of inter-plane tunneling strength and unconstrained $\mu$ phase space in pairing as well as study $q=(\pi,\pi)$ spin susceptibility using the same approach.

\section{Models and Methods}
\subsection{Non-interacting Hamiltonian }\label{sec2a}
We briefly summarize the tight-binding models for $\ell=2,3,$ and $4$.  This is necessary due to the large parameter space involved, which we discuss in detail below. 
\subsubsection{Bilayer Hubbard Model}
We consider a Bilayer ($\ell=2$) Hubbard Model with intra-planer nearest, $t$, and next-nearest, $t'$, single particle hopping as well as inter-planer tunneling between the two planes $t_{z}$ \cite{Hubbard1963}. Setting the lattice constant $a=1$, the non-interacting dispersion of the intra-layer component is given by
\begin{equation} \epsilon^{\parallel}(\textbf k)=-2t[\cos(k_x)+\cos(k_y)]-4t^\prime [\cos(k_x)\cos(k_y)].
\label{intra-disp}   
\end{equation}
  We incorporate a momentum-dependent inter-layer hybridization term $t_{\perp}(k)$ alongside an isotropic component $t_{bs}$ given  by 
\begin{equation}
  t_{z}(k) = t_{bs} + t_{\perp}[\cos(k_{x})-\cos(k_{y})]^2.\label{eq:tz}  
\end{equation}
This $t_{z}(k)$ hybridization term has been previously studied for different families of cuprate materials and is derived from tight-binding parameters fitted to LDA calculations and confirmed by experimental observation \cite{ANDERSEN19951573,Andersen, Chakravarty1993,Markiewicz}. We assume that the hybridization term $t_{z}(k)$ is the same for a general $\ell$ layered system and set the c-axis lattice constant as $c=1$ for simplification. This formulation leads to the construction of the complete non-interacting Hamiltonian $H_{0}$:
\begin{align}
    \nonumber H_{0}^{\ell=2} &= \sum_{k\sigma} \vec{c}^{\dagger}_{k\sigma} \hat{\xi}_{k} \vec{c}^{}_{k\sigma} \\
    H_{0}^{\ell=2} &= \sum_{k} \begin{bmatrix}
        c^{1\dagger}_{k\sigma} & c^{2\dagger}_{k\sigma}  
    \end{bmatrix}
    \begin{bmatrix}
        \epsilon_\parallel(k) & t_{z}(k) \\
        t_{z}(k) & \epsilon_\parallel(k)       
    \end{bmatrix}
    \begin{bmatrix}
        c^{1}_{k\sigma} \\
        c^{2}_{k\sigma}  
    \end{bmatrix},
\end{align}
where $c^{1\dagger}_{k\sigma}(c^{1}_{k\sigma})$ and $c^{2\dagger}_{k\sigma}(c^{2}_{k\sigma})$ represent the creation (annihilation) operators of spin $\sigma$ electron with momenta $k$ in plane index $i$ indicated by the superscript $1,2$. Diagonalizing $\hat{\xi}_{k}$ yields:
\begin{equation*}
  H_{0}^{\ell=2} = \sum_{k\sigma} \begin{bmatrix}
        a^{1\dagger}_{k\sigma} & a^{2\dagger}_{k\sigma}  
    \end{bmatrix}
    \begin{bmatrix}
        \tilde{\epsilon}^{1} (k) & 0 \\
        0 & \tilde{\epsilon}^{2}(k)       
    \end{bmatrix}
    \begin{bmatrix}
        a^{1}_{k\sigma} \\
        a^{2}_{k\sigma}  
    \end{bmatrix}  
\end{equation*}
where $a^{1\dagger}_{k\sigma}(a^{1}_{k\sigma})$ and $a^{2\dagger}_{k\sigma}(a^{2}_{k\sigma})$ represents the creation(annihilation) operator of spin $\sigma$ the bonding ($k_{z}=0$) and anti-bonding ($k_{z}=\pi$) momenta plane respectively. The $a^{1,2}_{k\sigma}$ eigenvector can be expressed as linear combination of uncorrelated basis $a^{1,2}_{k\sigma} =\frac{1}{\sqrt{2}} (c^{1}_{k\sigma} \pm c^{2}_{k\sigma})$. The eigenvalue of  $a^{1,2}_{k\sigma}$ represents the full dispersion with inter-planer hybridization given by  $\tilde{\epsilon}^{1,2}(\textbf k) = \epsilon^\parallel(k)  \pm t_{z}(k)- \mu_{1,2}$ where $\mu_{1,2}$ is the chemical potential that controls the filling in the respective bands. This reduces the bilayer Hamiltonian into an effective two-orbital problem where the $k_{z}$ index can treated as a separate band index. An isotropic bilayer band splitting would be obtained for $t_{\perp}=0$ and  $t_{bs}\neq 0$ case. Conversely, $t_{\perp}\neq 0$ and  $t_{bs}= 0$ would result in highly anisotropic splitting with a maximum gap of $4t_{\perp}$ at the anti-nodal point $k=[\pi,0]$  and dispersion-less in the c-axis along diagonal lines $|k_{x}| = |k_{y}|$.

\subsubsection{Trilayer and Quadlayer Hubbard model }
 Considering inter-planar single particle hopping between only the adjacent planes and utilizing the same intra-layer dispersion defined in Eq.~\eqref{intra-disp}, the trilayer $(\ell=3)$ Hamiltonian can be defined as 
\begin{align}
 H^{\ell =3}_{0}&= \sum_{k \sigma} \begin{bmatrix}
        c^{1\dagger}_{k\sigma} & c^{2\dagger}_{k\sigma}  & c^{3\dagger}_{k\sigma}  
    \end{bmatrix}
    \begin{bmatrix}
         \epsilon^{\parallel}(k) & t_{z}(k) & 0\\
        t_{z}(k) &  \epsilon^{\parallel}(k) & t_{z}(k)\\
         0 & t_{z}(k) & \epsilon^{\parallel}(k)\\
    \end{bmatrix}
    \begin{bmatrix}
        c^{1}_{k\sigma} \\
        c^{2}_{k\sigma}\\
        c^{3}_{k\sigma}
    \end{bmatrix}
\end{align}

Similar to the Bilayer case, the Hamiltonian above can be diagonalized on the basis of the eigenvectors. The $a^{1,3}_{k \sigma} = c^{1}_{k\sigma} \pm \sqrt{2} c^{2}_{k\sigma} + c^{3}_{k\sigma}$ eigenvectors yield a bonding $(k_{z}=0)$ and an anti-bonding $(k_{z}=\pi)$  band corresponding to the two outer plane in reciprocal space, given by  $\Tilde{\epsilon}^{1,3}_{k \sigma} = \epsilon_{\parallel}(k) \pm \sqrt{2} t_{z}(k) -\mu_{1,3}$. Meanwhile, the eigenvector $a^{2}_{k\sigma} = (c^{1}_{k\sigma} - c^{3}_{k\sigma})/ \sqrt{2}$ results in an inner planar dispersion $(k_{z}=\frac{\pi}{2})$ given by $\Tilde{\epsilon}^{2}_{k \sigma} = \epsilon{^\parallel}(k) - \mu_{2}$ that is independent of $t_{z}(k)$, thereby identical to the single-layer model. This parameterization allows us to treat $k_{z}=[0,\frac{\pi}{2},\pi]$  momenta labels as band indices where   $\Tilde{\epsilon}^{1}_{k \sigma}\leq \Tilde{\epsilon}^{2}_{k \sigma} \leq \Tilde{\epsilon}^{3}_{k \sigma}$, effectively recasting the trilayer model to a three-orbital problem. With the inclusion of an additional layer, the quad-layer $(\ell=4)$ Hamiltonian is constructed as
\begin{multline}
 H^{\ell=4}_{0}= \sum_{k \sigma} \begin{bmatrix}
        c^{1\dagger}_{k\sigma} & c^{2\dagger}_{k\sigma}  & c^{3\dagger}_{k\sigma} & c^{4\dagger}_{k\sigma} 
    \end{bmatrix}\\
    \begin{bmatrix}
        \epsilon^{\parallel}(k) & t_{z}(k) & 0 & 0\\
        t_{z}(k) &  \epsilon^{\parallel}(k) & t_{z}(k) & 0\\
         0 & t_{z}(k) &  \epsilon^{\parallel}(k) & t_{z}(k)\\
          0 &0  & t_{z}(k) & \epsilon^{\parallel}(k)\\
    \end{bmatrix}
    \begin{bmatrix}
        c^{1}_{k\sigma} \\
        c^{2}_{k\sigma}\\
        c^{3}_{k\sigma}\\
        c^{4}_{k\sigma}
    \end{bmatrix}.   
\end{multline}
The $\ell=4$ Hamiltonian is readily diagonalized to obtain four distinct dispersions given by $\Tilde{\epsilon}^{1,2}_{k \sigma}= \epsilon^{\parallel}(k) - t_{z}(k)(\sqrt{5} \pm 1)/2 -\mu_{1,2}$ and $\Tilde{\epsilon}^{3,4}_{k \sigma} = \epsilon^{\parallel}(k)  + t_{z}(k)(\sqrt{5} \pm 1)/2 - \mu_{3,4}$ parameterized by band indices  $k_{z}=[0,\frac{\pi}{3},\frac{2\pi}{3},\pi] $ such that $\Tilde{\epsilon}^{1}_{k \sigma}\leq \Tilde{\epsilon}^{2}_{k \sigma} \leq \Tilde{\epsilon}^{3}_{k \sigma} \leq \Tilde{\epsilon}^{4}_{k \sigma}$. 

\subsubsection{Tight binding fitting parameters}
The challenge in conducting a comparative study among $\ell=1,2,3,4$ models lies in the vast parameter space one needs to probe. This is exacerbated by the fact that LDA fits for tight-binding parameters in cuprates vary depending on the layer number and the family of homologous series it belongs to. This necessitates adopting the same tight-binding parameters across the $\ell$ layered models to facilitate a direct and clear comparison. First, energies are normalized in the unit of $t$ by setting $t=1$. Following existing literature on theoretical studies of bilayer and trilayer systems, we fix $t^\prime/t=-0.3$ and anisotropic inter-layer tunneling strengths within the range $0.05 \leq t_{\perp}/t \leq 0.150$ \cite{pairhopping(3), Bilayer(Gan), Trilayer(1), Trilayer(2), Bilayer(IWano), Bilayer(Zegrodnik)}. Here, we set the isotropic tunneling term $t_{bs}/t=0$ and primarily study the  $t_{\perp}/t=0.125$ anisotropic case for all $\ell$-layered systems. The results for non-zero isotropic tunneling $t_{bs}$ and the dependence of $t_{\perp}$ on pairing are provided in the supplemental materials.  We fix $\mu$ within each band to further reduce the parameter space.  For a fixed $\mu$, the presence of a non-zero $t_z$ leads to layer-differentiated doping in the non-interacting reciprocal layers, which is explored in detail in the supplemental
materials.

\subsection{Interacting Hamiltonian}
The interacting component of the Hamiltonian includes both intra-layer and inter-layer components, represented as follows:
\begin{multline}
    H_{U} =  \frac{U}{2}\sum_{i,\ell,\sigma \neq\sigma^\prime } n_{i\ell\sigma} n_{i\ell\sigma^{\prime}} + \frac{V}{2}\sum_{i,\ell\neq
\ell^\prime, \sigma} n_{i\ell\sigma} n_{i \ell^{\prime}\sigma'} \\ + J\sum_{i,\ell\neq
\ell^\prime,\sigma \neq \sigma^\prime  }c^\dagger_{i\ell\sigma }c^\dagger_{i\ell\sigma^\prime}c_{i\ell^\prime\sigma^\prime}c_{i \ell^\prime\sigma} \\+ J^\prime \sum_{i,\ell\neq
\ell^\prime,\sigma \neq \sigma^\prime  }c^\dagger_{i\ell\sigma }c^\dagger_{i  +\delta,\ell\sigma^\prime}c_{i+\delta,\ell^\prime\sigma^\prime}c_{i \ell^\prime\sigma}.
\end{multline}

Here, $n_{i \ell \sigma}$ denotes the number operator for electron with $\sigma$ spin in the $i$th site and $\ell$th layer and  $\delta \in \{\hat{-x},\hat{-y},\hat{x},\hat{y}\}$ refers to the nearest neighbor site. $U$ represents the onsite local interaction between electrons with opposite spins, and $V$ denotes the inter-layer interaction between two electrons in adjacent layers, separated by a distance of the c-axis lattice constant. We also incorporate inter-layer pair hopping of two configurations: local \emph{onsite} pair hopping $J$ and a nonlocal \emph{offsite} pair hopping $J^{\prime}$. In the offsite configuration, a pair of electrons with opposite spins sits on the nearest neighbor and scatters to the next layer without any spin flip. Both $J$ and $J^{\prime}$ could be considered as microscopic mechanisms behind Josephson coupling, but rather than being a second-order kinetic process, they are interpreted here as Coulombic processes \cite{pairhopping(1),pairhopping(2),pairhopping(3)}. This formulation allows for the scattering of a Cooper pair between the next adjacent and next-next adjacent momenta layers via two or more inter-layer pair hopping interactions in the $\ell=3$ and $\ell=4$ Hubbard model. The long-range Coulombic process $J^\prime$, when Fourier transformed into momentum space has a form factor $J^{\prime}(q) = 2(\cos(q_{x}) + \cos(q_{y}))$.\cite{pairhopping(3)} We incorporate the effect of $J'$ inside every $J$ interaction via parameterization of $J$ as $J(q) \rightarrow J(1 + 2J'/J [\cos(q_{x})+ \cos(q_{y})])$ which requires fixing the ratio $J'/J$ to a specific value. 

\subsection{Pair Correlation in Multi-layered Hubbard model}
To investigate the pairing potential in the normal state of the layered Hubbard model, we calculate the uniform ($\mathcal{\mathbf{Q}}=0$) and static ($\mathbf{\Omega} = 0$) pairing susceptibility from linear response theory \cite{Chen:2015,Rohringer12}
\begin{equation}
\chi_{g}= \frac{1}{\ell}\int^{\beta}_{0} d\tau \langle \Delta^{\dagger }_{g}(\tau) \Delta_{g}(0)\rangle e^{i\Omega \tau}, \label{eqn:pairing}
\end{equation}
where $\beta$ is the inverse temperature $1/T$, $g$ denotes the symmetry factor of the superconducting order parameter $\Delta^{\dagger}_{g} = \sum_{\mathbf{k}}g(k)c^{\dagger}_{\mathbf{k}\downarrow}c^{\dagger}_{\mathbf{-k}\uparrow}$, and $\tau$ represents the imaginary time. The summation over $k_{x}$, $k_{y}$, and $k_{z}$ inside the expectation value is implicitly implied. The pair correlation function is normalized by the number of layers ($\ell$). One can perform explicit summation over $k_{z}$ momenta dependency within the expectation value of the Eq.~\ref{eqn:pairing}, resulting in 
\begin{equation}
  \chi_{g}(\tau) = \frac{1}{\ell}\sum_{k_{z},k^\prime_{z}}\underbrace{\big\langle \Delta_{ g}^{\dagger k_z}( \tau) \Delta_{g}^{k^{\prime}_{z}}(0)\big\rangle}_{ \chi^{k_{z},k^{\prime}_{z}}_{g}}.   
\end{equation}
From this, we can distinguish the different channels of pairing $\chi^{k_{z},k^{\prime}_{z}}_{g}$ that accounts for the correlation between two time-ordered cooper pairs residing in $k_{z}$ and $k'_{z}$ momenta plane.

In the layered system three types of symmetry factor are of particular interest: the $d_{x^2-y^2}$-wave $(\cos k_{x} - \cos k_{y})$ representing intra-layer pairing, the $d^{(1)}_{z}$-wave $(\cos k_{z})$ for inter-layer pairing between adjacent layers, and the $d^{(2)}_{z}$-wave $(\cos 2k_{z})$ for inter-layer pairing between next adjacent layers relevant in the context of the trilayer and quadlayer model\cite{sigrist:1991,Nbultut,Bilayer(Zegrodnik),tsuei:2000,trilayer_cuprate}. We primarily focus on $d_{x^2-y^2}$-wave pairing; the results on  $d^{(1)}_{z}$- and $d^{(2)}_{z}$-wave are not shown but discussed briefly in a later section. Interpretation of $\chi_{g}$ is that a positive response would signal an enhancement of the anomalous green's function corresponding to the symmetry factor $g(k)$.
One can re-write the Eq.~\eqref{eqn:pairing} in the form of the Bethe-Salpeter equation for the particle-particle channel as 
\begin{align}
    \chi_{g} &= \frac{\Bar{\chi}_{g}}{1 - \Gamma\Bar{\chi}_{g}}, \\
    \Bar{\chi}_{g} &= \frac{1}{\ell}\int^{\beta}_{0} d\tau \langle g(k)^2 c_{k\downarrow}(\tau)c^{\dagger}_{k\downarrow}(0)\rangle \langle c_{-k\uparrow}(\tau)c^{\dagger}_{-k\uparrow}(0)\rangle,
\end{align}
where  uncorrelated susceptibility $\Bar{\chi}_{g}$ is fully dressed particle-particle bubble and $\Gamma_{g} $ is the vertex insertion \cite{Salpeter}. When the eigenvalue of the $\Gamma\Bar{\chi}_{g}$ reaches unity $\chi_{g}$ diverges, and second-order phase transition to a superconducting state for the symmetry factor $g$ is attained. Since divergence to $\chi_{g}$ is an attribute of the vertex $\Gamma_{g}$, the strength of the vertex component of the pair correlation, $P_{g}$, can be utilized as an indicator of superconducting instability \cite{Chen:2015,White}, given by
\begin{align}
     P_{g} = \chi_{g} - \bar{\chi}_{g}
\end{align}
We shall evaluate the expectation value of $P_{g}$ diagrammatically order-by-order in powers of $U$,$V$, and $J$. To proceed with the expansion, we first make a simplifying  assumption that the non-interacting propagator can be expressed in the diagonal eigenbasis of the Hamiltonian,
\begin{align}
G_{ab} = \frac{\delta_{ab}}{i\omega - \epsilon_{ab}},
\label{G0}
\end{align}
where $\delta_{ab}$ represents the Kronecker delta. In the next section, we will make use of this property to construct our multiband perturbative expansion symbolically.

\begin{figure*}
    \centering
    \includegraphics[width=1\textwidth]{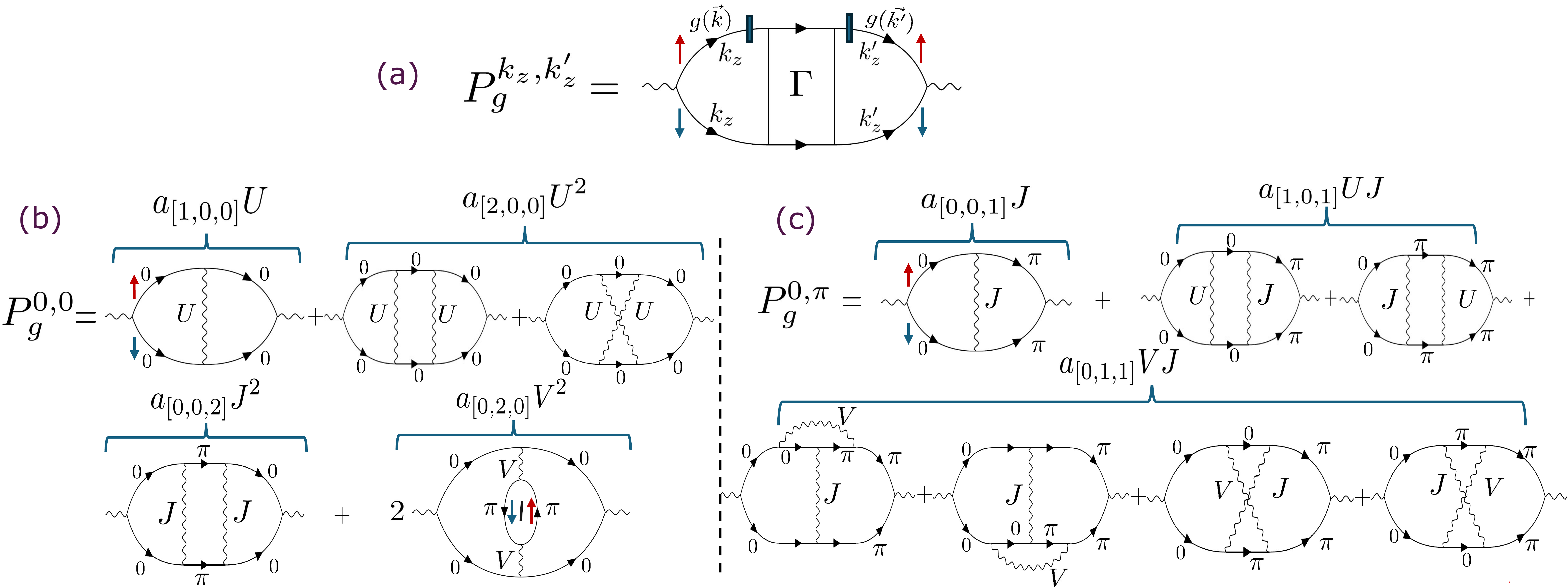}
    \caption{(a) Schematic illustrating the assignment of initial band indices $k_{z}$ and $k'_{z}$ to the particle-particle diagrams  corresponding to  pairing element $P^{k_{z}, k^{\prime}_{z}}_{g}$. The blue bar denotes the incoming and outgoing fermionic lines, where the symmetry factor $g(\vec{k})$ applies to the momenta. (b,c) Shows the particle-particle diagrams associated with first and second-order coefficients we generate in the $P^{0,0}_{g}$ and $P^{0,\pi}_{g}$ channel in $\ell=2$ system. $P^{\pi,\pi}_{d}$ channel has identical coefficient and diagram sets as $P^{0,0}_{d}$, which can obtained by substituting band index $0$ with $\pi$ and vice-versa. }
    \label{fig:schematics}
\end{figure*}

 \subsection{Diagrammatic Expansion in powers of $U-V-J$}
 To each of the matrix elements  $\chi^{k_{z},k^{\prime}_{z}}_{g}$ we apply  the `m'-th order correction of the interacting Hamiltonian
 \begin{equation}
     \chi^{k_{z},k^{\prime}_{z}}_{g} =\frac{(-1)^m}{m! \ell} \big\langle \mathcal{T} \Delta_{g}^{k_{z}}(\tau)\bigg[\int_{0}^{\beta}d\tau_{n} \prod^{m}_{n} H_{U}(\tau_{n}) \bigg] \Delta_{g}^{k'_{z}}(0) \big\rangle.
     \label{pertubative}
 \end{equation}
For any given order $m$, the expression yields the product of $2m+2$ creation and annihilation operators whose ensemble average must be evaluated. In a generalized multi-band problem, the standard procedure is to Wick contract the expression into a sum of all possible pairs, translating to a set of particle-particle Feynman diagrams from which all the disconnected and topologically indistinguishable diagrams are removed. The diagram sets are Fourier transformed and depend on a set of momentum and frequency-conserving labels. The Matsubara summation can be performed by repeated application of the residue theorem. The rest of the momentum dependency is integrated over stochastically via uniform distribution Monte Carlo sampling. While Matsubara sum can be done readily by hand for selected low-order topologies, the task becomes prohibitive at higher order due to the exponentially growing diagram set and complexity of the analytic expression. Therefore, we use the recently developed symbolic tool Algorithmic Matsubara integration (AMI), which automates the Matsubara frequency summation using residue theorem\cite{libami,AMI,torchami}. The symbolic AMI tool has found success in evaluating single-particle and two-particle correlation functions in single-band problems \cite{mcniven:2021,mcniven:2022, taheri:2020,burke:2023, Gazizova(2023),farid:2023}. 

The challenge in applying AMI to a generalized multi-band problem lies in generating a symbolic representation of Feynman diagrams with correct band indices and interactions assigned to them. This has been accomplished for a generalized multi-band model using a determinantal method that is suited to AMI.\cite{Assi} However, in the determinantal methods, it is impossible to separate the coefficients for series expansions based on $U$, $V$, and $J$ as the interactions become intertwined during the determinant construction. This necessitates the exhaustive computation of the entire diagram sets for each interaction parameter set. To circumvent this, we directly generate in advance all the distinct connected $\chi^{k_{z},k^{\prime}_{z}}_{g}$ particle-particle Feynman topologies, avoiding the determinant construction for Wick's decomposition altogether. We then exclude diagrams that are topologically indistinguishable, do not possess vertex components, or contain any Hartree insertions to filter out the required topologies to compute $P^{k_{z},k'_{z}}_{g}$ at fixed $\mu$. For every generated topology, if one were to naively assign every possible combination of band indexes to the propagators, it would produce numerous diagrams containing at least one or more interactions corresponding to a zero matrix element of the $H_{U}$ interaction. These diagram sets have no contribution, so sampling and computing them would result in a large, unnecessary computational cost. In order to avoid this, we introduce a new iterative scheme where we assign the band indexes based on only \emph{non-zero} matrix elements of interaction, ensuring that we directly generate and compute only the contributing diagram sets for $P^{k_{z},k'_{z}}_{g}$.

To do so, we first diagonalize the Hamiltonian to obtain the eigenbasis that forms our band indices. This allows us to construct the diagonal basis of our non-interacting propagator given by Eq.~\ref{G0}, ensuring only a single band indices can be assigned to each propagator. Similarly, the non-zero matrix elements of $H_{U}$ are parameterized on the basis of these band indices. For the matrix element of the pair correlation $P^{k_{z},k_{z}^\prime}_{g}$, we already know that two time-ordered order parameters belong to band indices $k_{z}$ and $k'_{z}$. Diagrammatically, this pertains to assigning $k_{z}$ and $k'_{z}$ to the two incoming fermionic lines and two outgoing lines, respectively, as shown by the schematics in Fig.~\ref{fig:schematics}(a). For the remaining $2m-2$ unassigned fermionic lines, we iteratively assign all possible combinations of band indices, with the selection criterion being that the interaction sets must correspond to non-zero matrix elements of the interaction. If an interaction not corresponding to $H_U$ emerges at any stage, the sampling process is immediately halted, and that particular combination of band indices is dismissed. We then proceed to the next possible combination. This process is repeated until the complete diagram set with correct band indices and non-zero interactions is generated for a given topology. 

This scheme can be automated for all topologies, quickly constructing the entire symbolic representation of the multiband diagrammatic expansion of $P^{k_{z},k^{\prime}_{z}}_{g}$  that is amenable to AMI. Furthermore, during the sampling process, one can easily book-keep the number of $U$,$V$, and $J$ interactions present in the diagram sets. Consequently, diagrams can be sorted and summed based on the number of each interaction present. This allows one to extract the coefficient of $P^{k_{z},k'_{z}}_{g}$ for a multi-power series expansion in the form
\begin{align}
    \label{eq:coeff}
    P^{k_{z},k'_{z}}_{g} &= \sum_{i,j,k} a_{[i,j,k]} U^{i}V^{j}J^{k} \nonumber \\
    &= a_{[1,0,0]}U + a_{[0,1,0]}V + a_{[0,0,1]}J + a_{[2,0,0]}U^2 \nonumber \\
    &\quad + a_{[0,2,0]}V^2 + a_{[0,0,2]}J^2 + a_{[1,1,0]}UV + a_{[1,0,1]}UJ \nonumber \\
    &\quad + a_{[0,1,1]}VJ + a_{[1,1,1]}UVJ + a_{[3,0,0]}U^3 + a_{[0,3,0]}V^3 \nonumber \\
    &\quad + a_{[0,0,3]}J^3 \ldots
\end{align}
Here  $a_{[i,j,k]}$ is a function of $k_{z}$ and $k'_{z}$ band indices that represent the coefficient for a particular combination of the powers of $U$, $V$, and $J$, as indicated by the indices $i$, $j$, and $k$ for the $m$th order, where $i+j+k=m$. From a single computation of non-zero coefficients, $P^{k_{z},k'_{z}}_{g}$ can be evaluated for any value of $U,V,$ and $J$ interaction strength. Moreover, these coefficients provide an understanding of the underlying microscopic mechanisms in which pairing occurs in a previously unexplored manner. 

The total correlated pairing susceptibility is computed as the sum of all vertex $P^{k_{z},k'_{z}}_{g}$ channels \cite{3band_Pd}  normalized by $\ell$
\begin{equation}
    P^{tot}_{g} = \frac{1}{\ell}\sum_{k_{z},k'_{z}}P^{k_{z},k'_{z}}_{g}.
\end{equation}

From the $P^{tot}_{g}$, we distinguish two different components of pairing: an intra-plane component  ($P^{\parallel}_{g}$) given by summing the diagonal channels $P^{\parallel}_{g} = \frac{1}{\ell}\sum_{k_{z}} P^{k_{z},k_{z}}_{g}$ and inter-plane component ($P^{\perp}_{g}$) that consists of a summation of all the off-diagonal channels  $P^{\perp}_{g} = \frac{1}{\ell}\sum_{k_{z} \neq k'_{z}} P^{k_{z},k'_{z}}_{g}$ with the total being $P^{tot}_{g}= P^{\parallel}_{g} + P^{\perp}_{g}.$ 

 Given the form of the interaction term, in order for $P^{\perp}_{g}$ to have a non-zero contribution in $P^{tot}_{g}$, a finite $J$ interaction is required. For a bilayer system ($\ell=2$) consisting of band indices $k_{z}=[0,\pi]$, one obtains
\begin{align}\label{eqstart}
 P^{\parallel}_{g}(\ell=2)= (P^{0,0}_{g} + P^{\pi,\pi}_{g})/2\\
 P^{\perp}_{g}(\ell=2) = (P^{0,\pi}_{g} + P^{\pi,0}_{g} )/2
\end{align}
and their summation
\begin{equation}
 P^{tot}_{g}(\ell=2) = (P^{0,0}_{g} + P^{\pi,\pi}_{g} +2P^{0,\pi}_{g})/2. 
 \label{eq:l2_pairing}
\end{equation}
Here we reduce the computation space of $P^{\perp}_{g}$  by taking advantage of the symmetry $P^{k_{z},k'_{z}}_{g} = P^{k'_{z},k_{z}}_{g}$. Extending this to trilayer $(\ell=3)$ case with band indices $k_{z}=[0,\frac{\pi}{2},\pi]$ , one obtains
\begin{align}
\label{eq:l2_parallel} P^{\parallel}_{g}(\ell=3)= (P^{0,0}_{g} +P^{\frac{\pi}{2},\frac{\pi}{2}}_{g}+ P^{\pi,\pi}_{g})/3\\
\label{eq:l2_perp}  P^{\perp}_{g}(\ell=3) = 2(P^{0,\frac{\pi}{2}}_{g} + P^{\frac{\pi}{2},\pi}_{g} + P^{0,\pi}_{g})/3
\end{align}
and their summation 
\begin{multline}
 P^{tot}_{g}(\ell=3) = (P^{0,0}_{g} + P^{\frac{\pi}{2},\frac{\pi}{2}}_{g}+P^{\pi,\pi}_{g})/3 \\+ 2(P^{0,\frac{\pi}{2}}_{g} + P^{\frac{\pi}{2},\pi}_{g}+ P^{0,\pi}_{g})/3.  
 \label{eq:l3_pairing}
\end{multline}

Eqs.~(\ref{eqstart})$\to$(\ref{eq:l3_pairing}) lead us to see that in any $\ell$-layered system, there are a total $(\ell)^{2}$  pairing channels of which there are $\ell$  diagonal $P^{k_{z},k_{z}}_{g}$ channels that contribute to $P^{\parallel}_{g}$ component and $\ell(\ell-1)$ off-diagonal $P^{k_{z},k'_{z}}_{g}$  channels that contribute to $P^{\perp}_{g}$ component. Upon summation and normalization by $1/\ell$, each of $P^{k_{z},k_{z}}_{g}$ channels inside $P^{\parallel}_{g}$  are averaged out while the  $P^{k_{z},k'_{z}}_{g}$  channels collectively contribute a net additive effect in $P^{\perp}_{d}$ with increasing layers. Thus, $P^{\parallel}_{g}$ can be regarded as intensive and $P^{\perp}_{g}$ as extensive property with respect number of layers.
 
We summarize the number of Feynman diagrams belonging to each of the inequivalent pairing channels and $P^{tot}_{g}$ for $m$ th order in Tab.~\ref{table} and Tab.~\ref{table1} for the $\ell=2$ and $\ell=3$ system. Note that the integral dimensionality in two-dimensional $(d=2)$ stacked lattice scales as $dm+d$. Here, we would like to highlight the enormity of the computational expense necessary to compute $P^{tot}_{g}$. For $\ell=2$, in the third order $P^{tot}_{g}$, one needs to perform eight-dimensional momentum integrals over the 276 particle-particle vertex diagrams for each data point. This complexity balloons to 3558 diagrams with an additional two integrals when progressing to the fourth order. This is further exacerbated by the fact that the $\ell=3$ system possesses more diagrams than the $\ell=2$ system due to increased spin-orbital indices and interaction matrix size.
 \begin{table}
    \raggedright
    {\large \textbf{$\ell=2$}}\\[4pt]
    \begin{tabular*}{1\linewidth}{@{\extracolsep{\fill}}|c|c|c|c|c|}
    \hline
     \hspace{1mm} m \hspace{1mm} & $P^{0,0}_{g}$ & $P^{0,\pi}_{g}$& $P^{\pi,\pi}_{g}$ & $P^{tot}_{g}$  \\
    \hline
    $1$ & 1 & 1  & 1 & 4 \\
    \hline
    $2$& 5 & 6  & 5 & 22 \\ 
    \hline
    $3$& 68 & 70  & 68 & 276 \\ 
    \hline
    $4$& 869 & 910  & 869 & 3558\\
     \hline
    \end{tabular*}
    \caption{\label{table}Table summarizing the total number of non-zero Feynman diagrams for the $\ell=2$ system at each respective order in the inequivalent pairing channels $P^{k_{z},k'_{z}}_{g}$ and the total sum $P^{tot}_{g}$ including the equivalent channels. }
\end{table}

 \begin{table}
    \raggedright
    {\large \textbf{$\ell=3$}}\\[4pt]
    \begin{tabular*}{1\linewidth}{@{\extracolsep{\fill}}|c|c|c|c|c|c|}
    \hline
    \hspace{1mm} m \hspace{1mm} & $P^{0,0}_{g}$/ $P^{\pi,\pi}_{g}$&  $P^{\frac{\pi}{2},\frac{\pi}{2}}_{g}$& $P^{0,\frac{\pi}{2}}_{g}$/ $P^{\frac{\pi}{2},\pi}_{g}$   &  $P^{0,\pi}_{g}$ &  $P^{tot}_{g}$  \\
    \hline
    $1$ & 1 & 1  & 1 & 0&  6\\
    \hline
    $2$& 5 & 8  & 6 & 1 & 39 \\ 
    \hline
    $3$& 68 &  123 & 79 & 11 & 529 \\ 
    \hline
    $4$& 904 & 1820  & 1100 & 177 & 7478\\
     \hline
    \end{tabular*}
    \caption{\label{table1}Table summarizing the total number of non-zero Feynman diagrams for the $\ell=3$ system at each respective order in the inequivalent pairing channels $P^{k_{z},k'_{z}}_{g}$ and the total sum $P^{tot}_{g}$ including the equivalent channels. }
\end{table}

\section{Result and Discussion} 
In this section, we study the total correlated susceptibility $P^{tot}_{d}$ along with its intra-plane $P^{\parallel}_{d}$ and inter-plane $P^{\perp}_{d}$ components in the $d_{x^2-y^2}$ ($d$) symmetry channel for $\ell=2,3$ and $4$ systems.  We conduct a detailed analysis of the $a_{[i,j,k]}$ coefficients for the different channels, enabling us to identify the diagram sets that enhance or suppress two pairing components over the \emph{U-V-J} interaction space. We also construct a generalized equation that speculates how  $P^{tot}_{d}(\ell)$ changes as a function of layer beyond $\ell =4$.  We primarily truncate our expansion at the third order and operate at high temperature and weak coupling limit by setting $\beta=5$ and  $U \leq 3.0$ with $V, J<U$. While temperature plays a key role, we restrict this parameter choice to ensure our perturbative expansion remains controlled, as demonstrated in our prior single-band 2D Hubbard model study \cite{farid:2023}. We examine the effect of fourth-order corrections on selected data sets due to prohibitive computational expense. Guided by previous studies, we fix the ratio of $J'/J =-0.5$, such that Cooper pairs remain in phase between the layers \cite{pairhopping(3),Bilayer(Zegrodnik)}. Here, we only study the anisotropic tunneling case with $t_{\perp}=0.125$ and $t_{bs}=0$. We study the effect of both isotropic and anisotropic tunneling in the supplemental materials. We find that inter-plane tunneling strength adversely affects $P^{tot}_{d}$, and there is no note-worthy difference between $t_{bs}$ and $t_{\perp}$. Several theoretical works, using different approaches, have studied single-particle and two particles properties in the layered system with both isotropic and anisotropic tunneling \cite{isotropic(scalepino),iso(Karakuzu),iso(kato),iso(Caplan),iso(Bouadim),Iso_1, Iso_2,pairhopping(3), Bilayer(Gan), Trilayer(1), Trilayer(2), Bilayer(IWano), Bilayer(Zegrodnik),Aniso}.
\subsection{Bilayer result}
\begin{figure}
    \centering
    \includegraphics[width=1\linewidth]{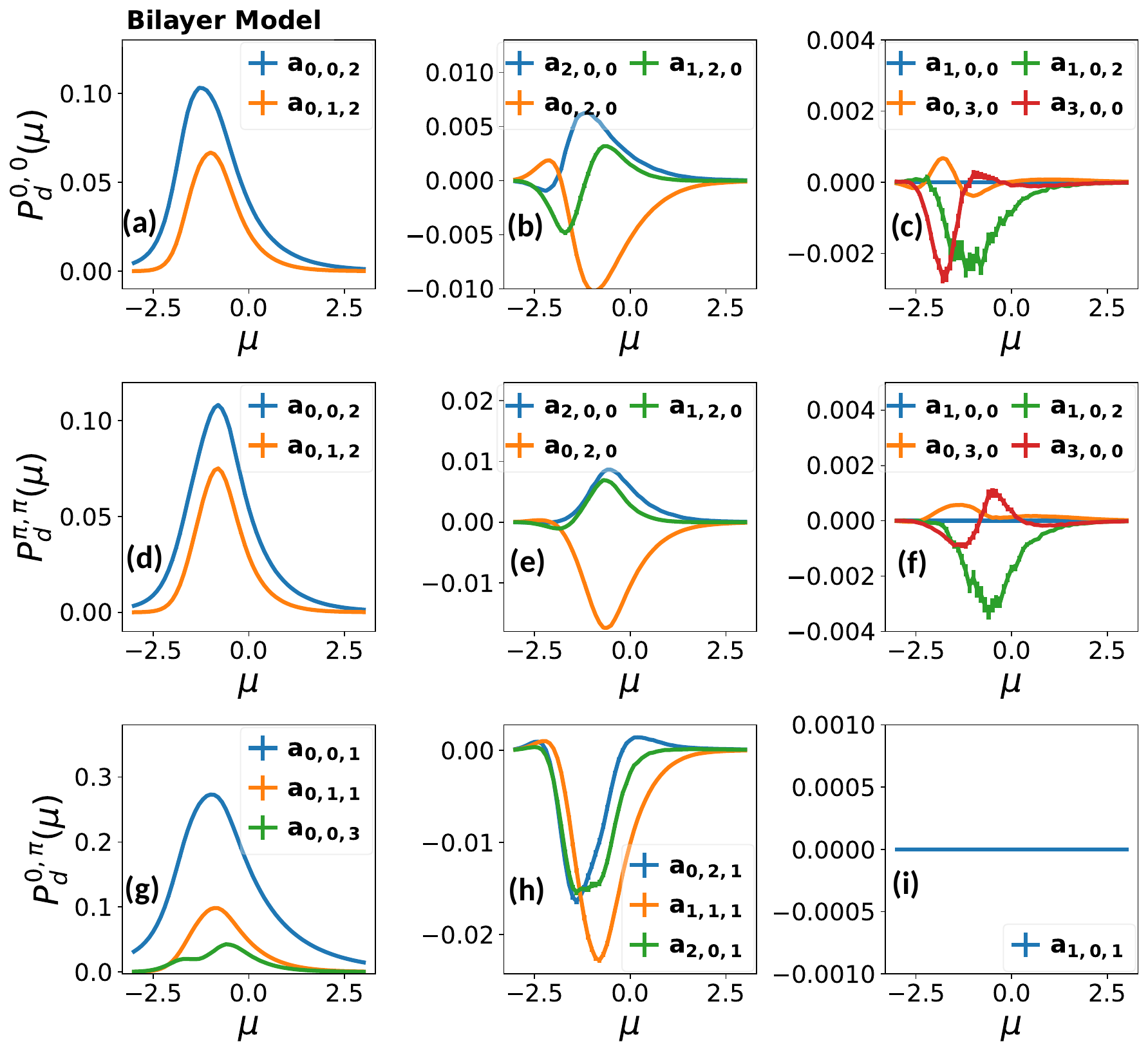}
    \caption{\label{coeff_bilayer} The $a_{[i,j,k]}$ coefficients of  
 $P^{k_{z},k^{\prime}_{z}}_{d}$ as a function of $\mu$ in the $\ell=2$ model determined by grouping particle-particle Feynman diagrams based on the number of $U$,$V$, and $J$ interactions present. The first row (a-c) shows intra-layer pairing in the bonding plane $P^{0,0}_d$, the second row (d-f) in the anti-bonding plane, and the third row (g-h) the off-diagonal $P^{0,\pi}_d$ components, representing inter-plane scattering. Here, we set $U,V,J =1$ with $J'/J=-0.5$.
 }
 \end{figure}

We present the  $a_{[i,j,k]}$ coefficients of the $P^{0,0}_{d}$,$P^{\pi,\pi}_{d}$ and $P^{0,\pi}_{d}$  channels for $\ell=2 $ system as a function of $\mu$ for $U,V,J=1$ up to third order in Fig.~\ref{coeff_bilayer}. The coefficients are arranged in columns based on their relative magnitude, from left (largest) to right (smallest). The schematic in Fig.~\ref{fig:schematics}(b,c) depicts all the first and second-order diagram sets corresponding to the non-zero coefficients for the $P^{0,0}_{d}$ and $P^{0,\pi}_{d}$ channels. It's worth noting that the $P^{\pi,\pi}_{d}$ channel shares identical sets of coefficients and corresponding diagrams as the $P^{0,0}_{d}$ channel. We remind the readers that since the offsite pair hopping term $J'$ is incorporated inside the onsite $J$ terms with the parameterization $J(q) \rightarrow J[1 + 2J'/J (\cos(q_{x})+ \cos(q_{y}))]$, the value of $J'$ is fixed to the ratio $J'=-0.5J$. The  $\mu$ in the $k_{z}=[0,\pi]$ bands is fixed such that doping in the two bands is equal. Multi-power series expansion is performed  to  $a_{[i,j,k]}$ coefficients using the  Eq.\ref{eq:coeff} to evaluate each of the $P^{k_{z},k'_{z}}_{d}$ channel in the \emph{U-V-J} interaction space. These coefficients will be utilized later to identify diagram sets that enhance or reduce pairing in the two components.

\begin{figure}
    \centering
    \includegraphics[width=1\linewidth]{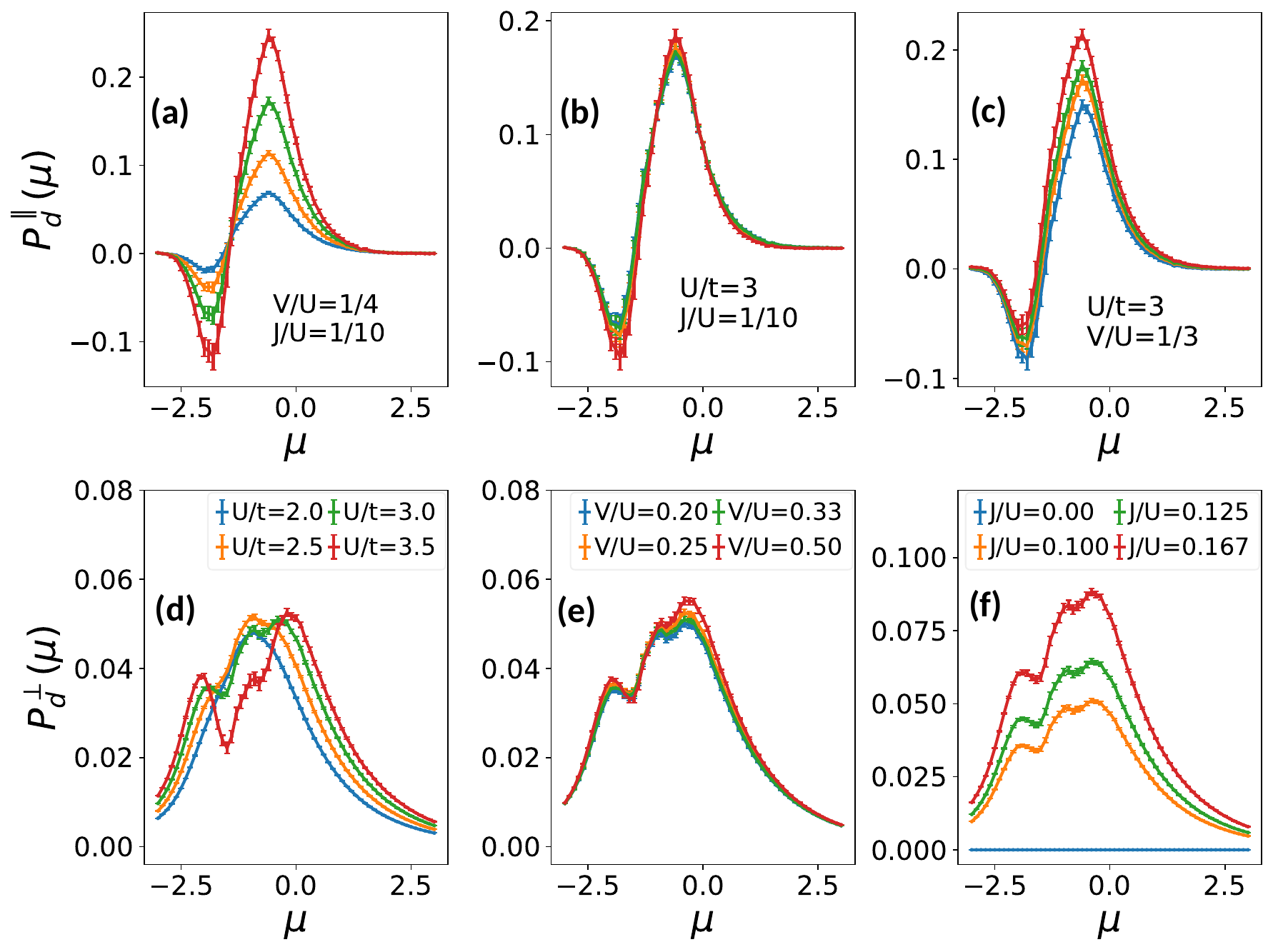}
    \caption{\label{comp_bilayer}  $P^{\parallel}_{d}$ and $P^{\perp}_{d}$ component obtained for $\ell=2$ system by performing   multi-power-series expansion on the coefficients in Fig.~\ref{coeff_bilayer} is  plotted as functions of the chemical potential $\mu$. Panels (a,d) vary $U/t$ with fixed $V/U=0.25$ and $J/U=0.10$, panels (b,e) vary $V/U$ with fixed $U/t=3.0$ and $J/U=0.10$ and panels (c,f) vary $J/U$ with fixed $U/t=3.0$ and $V/U=0.25$.   
 }
 \end{figure}

 Using the  Eq.~\ref{eq:l2_pairing}, Eq.~\ref{eq:l2_parallel} and Eq.~\ref{eq:l2_perp} on the three channels of pairing, one can obtain two pairing components $P^{\parallel}_{d}$ and $P^{\perp}_{d}$ and their sum, $P^{tot}_{d}$. In order to understand the role interaction in $P^{tot}_{d}$ pairing, we first plot truncated third order $P^{\parallel}_{d}$ and $P^{\perp}_{d}$ as a function of $\mu$ in Fig.~\ref{comp_bilayer}. We systematically probe the interaction space in the three panels by discretely varying one of $U$, $V$, and $J$ while keeping the others fixed. Firstly, it is observed that the peak structure resides in the negative $\mu$ region. 
 In Fig.~\ref{comp_bilayer}(a,d), when we keep $V/U=0.25$ and $J/U=0.10$ fixed, we observe that as $U$ increases, the pairing amplitude in $P^{\parallel}_{d}$ increases, while in $P^{\perp}_{d}$ it remains relatively unchanged. However, $P^{\perp}_{d}$ shows a shift in peak structure and location, with the formation of multiple valleys encompassing the $\mu=0$ to $\mu=-2$ region, especially noticeable in the cases of $U=3.0$ and $U=3.5$. It's important to note that the $U$ dependent features are masked by the fixed ratio $J/U$, as increasing $U$ also increases $J$, which affects $P^{\perp}_{d}$. When the absolute value of $J$ is fixed, increasing $U$ weakens the $P^{\perp}_{d}$ component. In Fig.~\ref{comp_bilayer}(b,e), we examine the $V/U$ dependence of the pairing component with fixed $U=3$ and $J/U=0.10$. The strength and peak structures of $P^{\parallel}_{d}$ and $P^{\perp}_{d}$ are largely unaffected for the four choices of $V$. Finally, in Fig.~\ref{comp_bilayer}(c,f), we increase $J/U$ while keeping $U=3$ and $V/U=0.25$ fixed and notice an overall enhancement in both $P^{\parallel}_{d}$ and $P^{\perp}_{d}$ component. While $P^{\perp}_{d}$ has no contribution at $J=0$, a significantly larger enhancement is observed with increasing $J$ compared to the $P^{\parallel}_{d}$ case.
 
\begin{figure}
    \centering
   \includegraphics[width=1\linewidth]{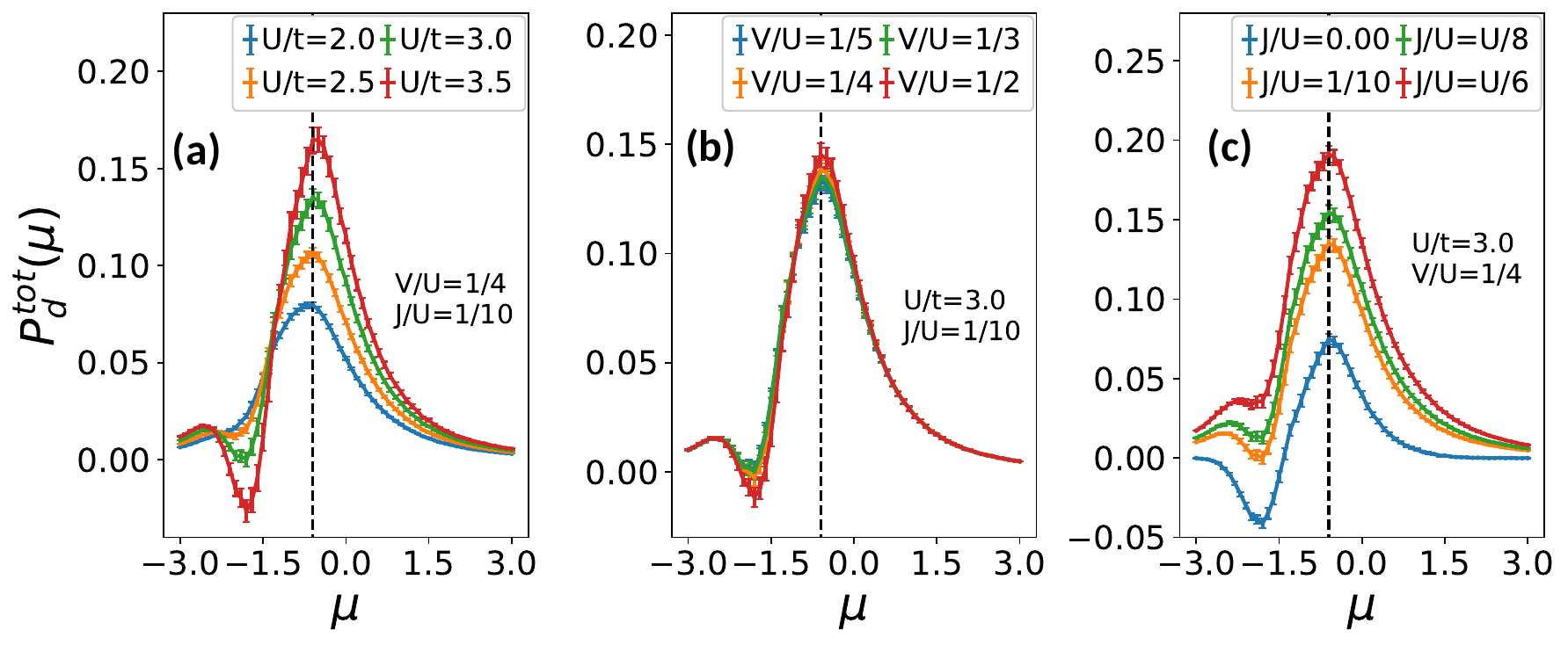}\\%
    \vspace{0.05cm}\includegraphics[width=1\linewidth]{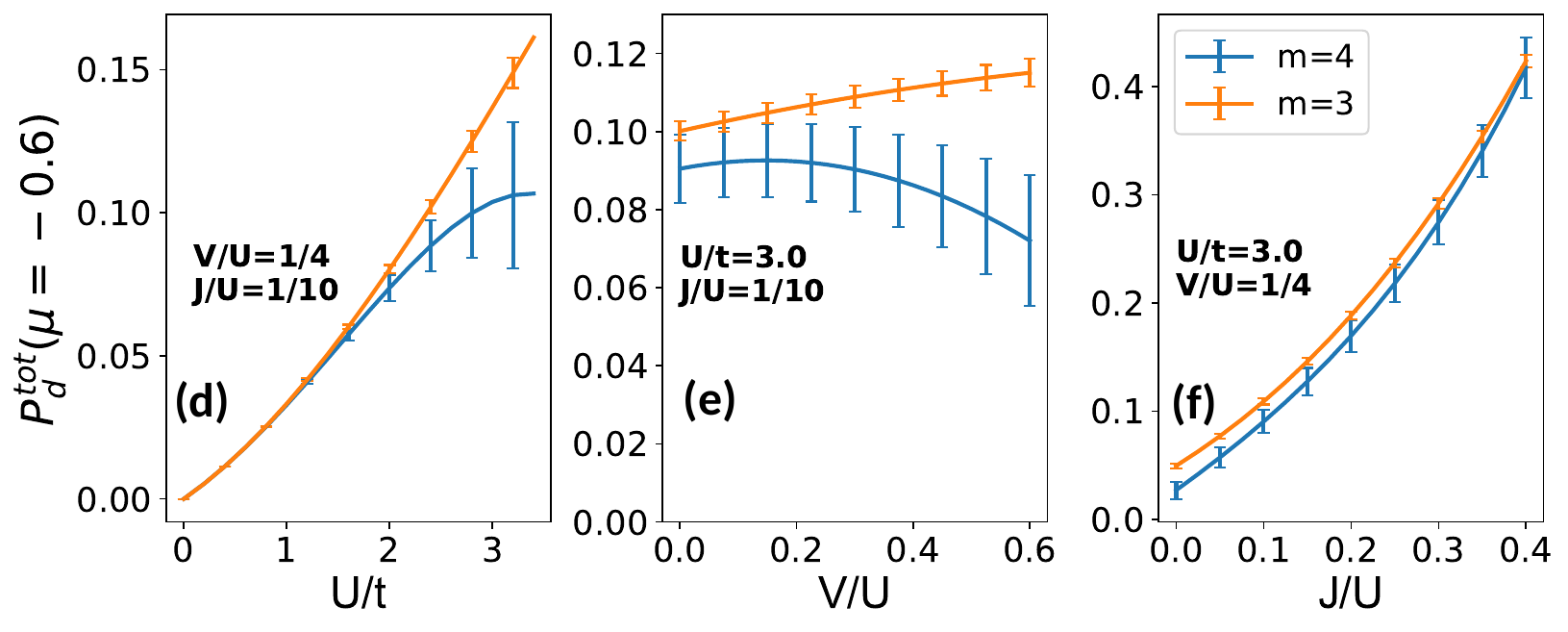}
    \caption{\label{UVJ_Ptot} $P^{tot}_{d}$ in $\ell=2$ system obtained by summing the $P^{\perp}_{d}$ and $P^{\parallel}_{d}$ components in Fig.~\ref{comp_bilayer} plotted as function of $\mu$ (a-c). The blacked dotted line at $\mu=-0.6$ shows where truncated 3rd order $(m=3)$ $P^{tot}_{d}$ is peaked for $U/t=3$ case. Fourth-order correction $(m=4)$ is applied to $P^{tot}_{d}$ and plotted as a function of (d) $U/t$, (e)$V/U$ and (f) $J/U$ at $\mu=-0.6$.  } 
\end{figure}

 The cumulative effect of $P^{\parallel}_{d}$ and $P^{\perp}_{d}$ in interaction space is studied by plotting $P^{tot}_{d}$  as a function of $\mu$ in Fig.~\ref{UVJ_Ptot}(a-c), while maintaining the same parameter selection as in Fig.~\ref{comp_bilayer}. It is observed that $P^{tot}_{d}$ shows a peak structure centered around $\mu=-0.6$ in all cases. The strength of the peak is positively correlated to the values of $U$ and $J$, where the former mainly originates from the $P^{\parallel}_{d}$ component and the latter is influenced by both $P^{\parallel}_{d}$ and $P^{\perp}_{d}$ components.

 Given the enormity of the diagram space presented in Tab.~\ref{table}, we resort to computing the fourth order for only a single data point. There are no conceptual hurdles evaluating beyond the fourth order but a computational difficulty associated with evaluating many Feynman diagrams with growing integral dimensionality that scales as $2m+2$. We apply the fourth-order correction at $\mu=-0.6$, where the third-order $P^{tot}_{d}$ is at its peak, and examine it as a continuous function of $U, V$, and $J$ in Fig.~\ref{UVJ_Ptot}(d-f). In Fig.~\ref{UVJ_Ptot}(d), we notice that $P^{tot}_{d}$ increases with increasing $U$, with the onset of saturation only appearing for $U > 3$ in the fourth-order case. This follows the pattern observed in the single-layer case, where the fourth and fifth-order terms have negative contributions that suppress pairing \cite{farid:2023}. On the other hand, increasing $V$ has a contrasting effect in Fig.~\ref{UVJ_Ptot}(e), where the truncated third order slightly enhances $P^{tot}_{d}$, while the fourth-order correction suppresses it at the large $V$ limit.  In Fig.~\ref{UVJ_Ptot}(f), we find that enhancement in $P^{tot}_{d}$  with $J$ exhibits a quadratic scaling with a prominent linear component for both the truncated third and fourth-order expansions. This pattern can be attributed to the quadratic scaling of $P^{\parallel}_{d}$ and the linear scaling of $P^{\perp}_{d}$ with respect to $J$ (not shown), indicating a substantial contribution primarily from first- and then second-order diagrams. This explains why the fourth-order correction has a negligible impact even when a large $J/U=0.4$ ratio is used.

\subsection{Analysis of coefficients in $\ell=2$}
To dissect the microscopic mechanism driving bilayer pairing, we now take a closer look at $a_{[i,j,k]}$ of all the inequivalent $P^{k_{z},k'_{z}}_{d}$ channels that are present in Fig.~\ref{coeff_bilayer}. Identifying the dominant coefficient makes it possible to pinpoint the key particle-particle diagrams and the combinations of $U, V$, and $J$ that positively contribute to pairing. Since we utilize a $U$ interaction that is significantly higher than $V$ and $J$, we first focus on the dominant coefficients consisting of only $U$ interactions given by $a_{[i,0,0]}$ coefficients with $V,J=0$. For  $P^{0,0}_{d}$ and $P^{\pi,\pi}_{d}$ case, the $a_{[1,0,0]}$ coefficient consisting of a first-order ladder diagram has no contribution. This is because the factorizability of ladder diagrams makes the symmetry projection of $d_{x^2-y^2}$ independent of each other, resulting in a zero contribution for $\mathcal{\mathbf{Q}}=0$ pairing. The largest contribution originates from the second-order $a_{[2,0,0]}$ coefficient consisting of a ladder and a crossed-interaction line topology shown in Fig.~\ref{fig:schematics}(b) and followed by $a_{[3,0,0]}$ coefficient consisting of the same 3rd-order diagrams set that appears in single layer Hubbard model case \cite{farid:2023}. Summing the $a_{[i,0,0]}$ coefficients for respective $P^{k_{z},k_{z}}_{d}$ diagonal channels would represent the single band case, and the results are consistent with our previous single-layer study. While $a_{[2,0,0]}$ and $a_{[3,0,0]}$ coefficients are relatively smaller, they constitute by far the biggest contribution when expanded in powers of $U>2$.

Turning our attention to the dominant raw coefficients at $U,V,J=1$, off-diagonal $P^{0,\pi}_{d}$ channel includes leading coefficients $a_{[0,0,1]}$  represented by a single first-order ladder diagram followed by $a_{[0,1,1]}$, whose diagram sets consist of a single pair hopping interaction $J$ as shown in Fig.~\ref{fig:schematics}(c). The physical interpretation of these coefficients is the pairing contribution that originates from a Cooper pair hopping from one plane to the adjacent plane via a single $J$ interaction. In addition, the diagonal $P^{0,0}_{d}$ and $P^{\pi,\pi}_{d}$ channels have $a_{[0,0,2]}$ as the leading coefficient, which consists of a single second-order ladder diagram, and $a_{[0,1,2]}$ as subdominant coefficient with diagrams sets also involving two $J$ interactions. These coefficients indicate pairing contribution from a Cooper pair hopping to the nearest plane and then returning to the starting plane via two $J$ interactions.

When comparing all the coefficients presented above, $a_{[0,0,1]}$ coefficient in the off-diagonal $P^{0,\pi}_{d}$ channel stands out as the largest coefficient that is roughly three times greater than the next leading coefficient $a_{[0,0,2]}$ in $P^{\pi,\pi}_{d}$ diagonal channel. Unlike $a_{[i,0,0]}$ coefficients expanded in $U/t=3$, we employ a $J$ value less than unity, indicating that higher-order coefficients involving only $J$ will make increasingly lesser contributions. As a result, lower-order coefficients make a greater contribution to $P^{tot}_{d}$ when expanded in powers of $J$. This suggests that the rapid growth of $P^{tot}_{d}$ with increasing $J$  primarily originates from the  $P^{\perp}_{d}$ component that is linear with $J$ and a secondary contribution from the $P^{\parallel}_{d}$ component that scales quadratically $(J^2)$.

Now, a contrasting picture emerges where ladder diagrams involving $U$ in $a_{[i,0,0]}$ coefficients having zero contribution while those with $a_{[0,0,j]}$ have the largest contribution. This is a direct consequence of the incorporation of  $J'$ with $J$ where $2(\cos(q_{x}) + \cos(q_{y}))$ form factor is applied to momentum transfer, $\mathbf{q}$, associated with $J$ vertex correction in ladder diagrams. As a result, the ladder diagrams are no longer factorizable and yield a significant contribution to $P^{tot}_{d}$, particularly the first-order ladder diagram. The contribution of these ladder diagrams is pinned to the ratio and sign of $J'/J$  employed. In the $J'/J=0$ case, the ladder diagrams mimic the factorizable nature of $a_{[i,0,0]}$ ladder diagrams, leading to zero contribution. The non-ladder diagrams results in marginal  enhancement to both $P^{\parallel}_{d}$ and $P^{\perp}_{d}$ components. With a positive $J'/J=0.5$, the signs of ladders diagrams inside $a_{[0,0,j]}$ for every odd $j$ is flipped, in particular the leading $a_{[0,0,1]}$ coefficient in $P^{0,\pi}_{d}$, driving $P^{\perp}_{d}$ repulsive with increasing $J$. Moreover, the magnitude of $J'/J$  utilized alters the extent of enhancement or suppression experienced. Therefore, we argue that $J'/J <0$ is a key requirement to increase pairing in multilayer systems over that of a single-layer system. Since these ladder diagrams are not part of spin and charge diagram sets, $J$ dependency of $P^{tot}_{d}$ will not be reflected on spin and charge susceptibility. Therefore, $q=(\pi,\pi)$ antiferromagnetic spin fluctuation that mediates the formation of $d$-wave anomalous greens self-energy is not the mechanism behind the enhancement of $P^{\perp}_{d}$ with $J$ \cite{Dong}. To demonstrate this, we have calculated the total intra-layer staggered spin susceptibility on the bilayer system in the supplemental materials.

In the previous section, we have discussed that increasing the value of $U$ has a detrimental effect on $P^{\perp}_{d}$ when the values of $J$ and $V$ are fixed. But the pairing amplitude remains relatively unchanged when the ratio of $J/U$ and $V/U$ is fixed, as shown in Fig.~\ref{comp_bilayer}(d). This can be explained by inspecting the $a_{[1,1,1]}$ and $a_{[2,0,1]}$ coefficients in Fig.~\ref{coeff_bilayer}(h) with negative amplitude in the $P^{0,\pi}_{d}$ channel. When expanded in powers of $U$, $V$, and $J$ interactions, they form a comparable negative contribution that competes with the positive influence of $a_{[0,0,1]}$ and $a_{[0,1,1]}$ coefficients. These opposing effects cancel each other out, resulting in a comparable $P^{\perp}_{d}$ amplitude.

\subsection{Trilayer Result}
In the bilayer case, our analysis has revealed that $a_{[0,0,2]}$ and $a_{[0,1,2]}$ are the two leading dominant coefficients in the diagonal channel. These coefficients are responsible for enhancing $P^{\parallel}_{d}$ with increasing  $J$. While coefficients $a_{[2,0,0]}$ and $a_{[3,0,0]}$  are small, they become the leading contributor to $P^{\parallel}_{d}$ when expanded with $U\geq2$. Furthermore, in the off-diagonal channel for $P^{\perp}_{d}$, the essential features are captured by first-order $a_{[0,0,1]}$ followed by second-order $a_{[0,1,1]}$ coefficients when restricted to small values of $V, J<1$.

In order to understand the pairing process in the trilayer model, we will first compare the dominant coefficients $a_{[i,j,k]}$ in $\ell=3$ with the coefficients in $\ell=2$ as a function of $\mu$. The raw coefficient for $\ell=3$ with $U,V, J=1$, and $J'/J=-0.5$ across all the inequivalent pairing channels $P^{k_{z},k'_{z}}_{g}$ is provided in the supplemental materials. One can qualitatively predict the pairing in the trilayer system by comparing the dominant coefficients in $\ell=3$ with the $\ell=2$.

\begin{figure}
    \centering
    \includegraphics[width=1\linewidth]{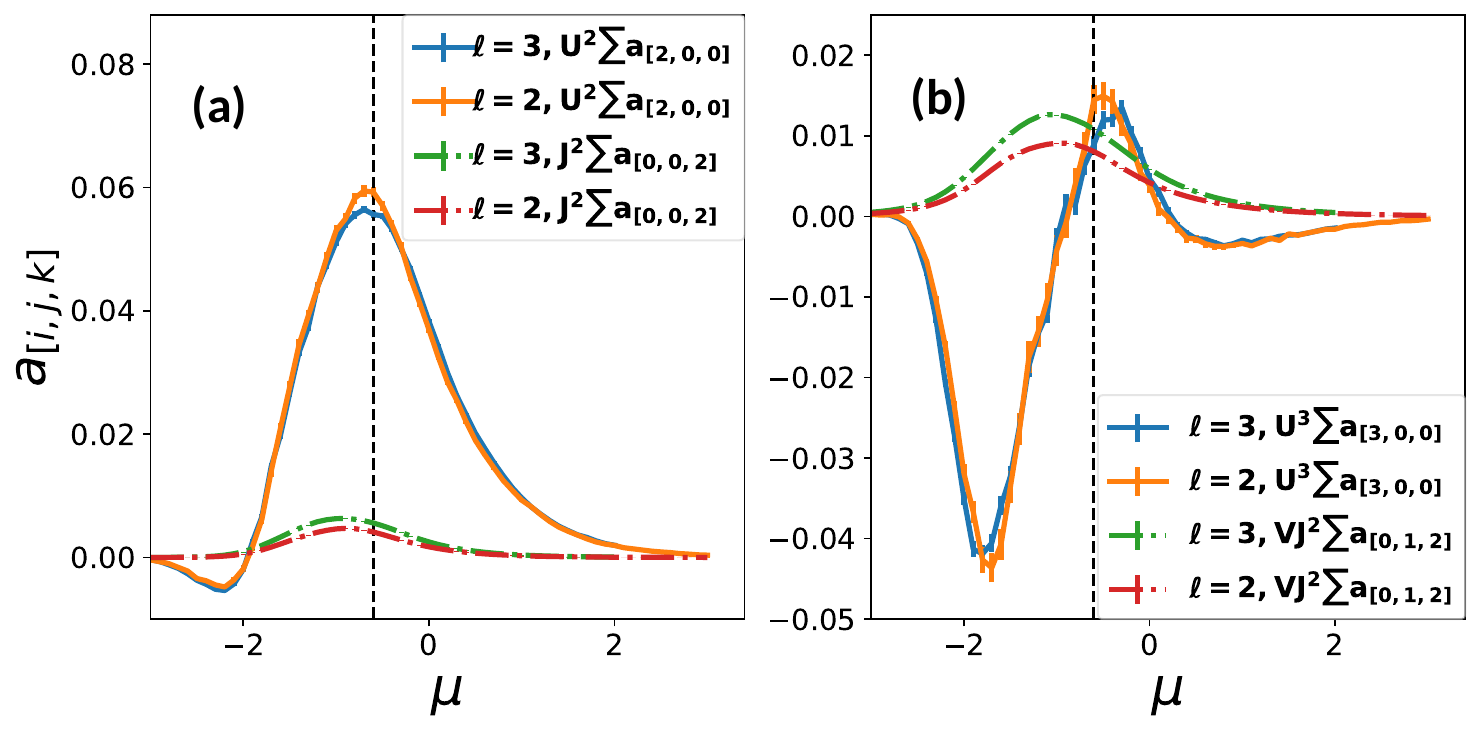}\\%
      \vspace{-0.15cm}\includegraphics[width=0.9\linewidth]{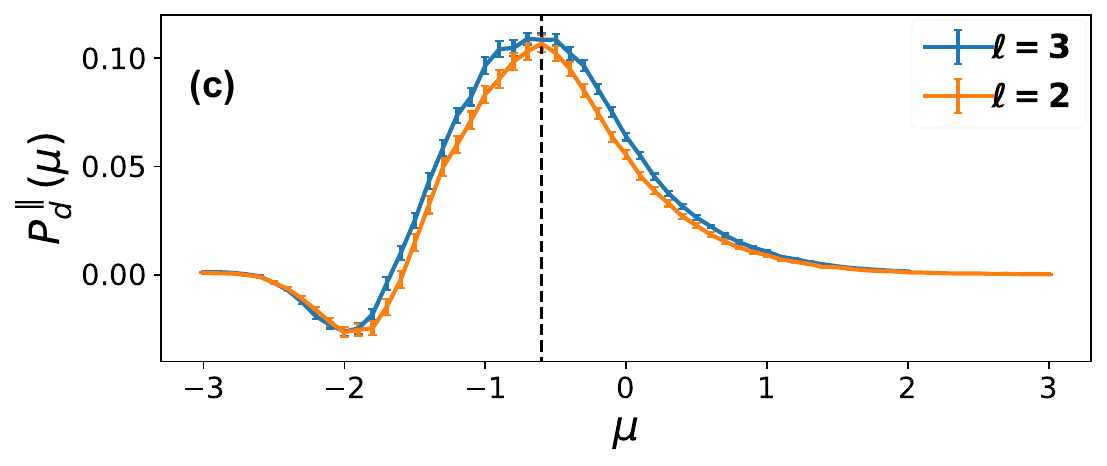}
    \caption{The overall contribution of the dominant coefficients in $\ell=2,3$ to $P^{\parallel}_{d}$ for $U=3,V=0.75,J=0.3$  as function of $\mu$ for (a) $a_{[0,0,2]}$ with dashed line and $a_{[2,0,0]}$ with solid line (b) $a_{[0,1,2]}$ with dashed line and  $a_{3,0,0}$ with solid line. (c) Comparison of $P^{\parallel}_{d}$ between $\ell=2$ and $\ell=3$ as function of $\mu$ with same choice of interaction strength. }
    \label{fig:para_coeff}
\end{figure}

Upon examining the $\ell=3$ coefficients in the $P^{0,0}_{d}$, $P^{\frac{\pi}{2},\frac{\pi}{2}}_{d}$, and $P^{\pi,\pi}_{d}$ diagonal channels, it is found that the non-zero coefficient sets are shared, including the dominant coefficient with those in the $P^{0,0}_{d}$ and $P^{\pi,\pi}_{d}$ channels for $\ell=2$ as shown in Fig.~\ref{coeff_bilayer}(a-f). Although there are slight variations in magnitude and peak locations due to the relative position of the van Hove singularity to the Fermi surface (see supplemental) when these channels are summed and averaged over $\ell$, the resulting $P^{\parallel}_{d}$ shows no significant differences. To illustrate this, we plot the overall contribution of leading coefficients $a_{[2,0,0]}$, $a_{[3,0,0]}$, $a_{[0,0,2]}$, and $a_{[0,1,2]}$ to $P^{\parallel}_{d}$ at $U=3.0$, $V=0.75$ and $J=0.3$ for the $\ell=2,3$ system in Fig.~\ref{fig:para_coeff}(a,b). It is noted that the positive peak of $a_{[2,0,0]}$ and $a_{[3,0,0]}$ in $\ell=3$ is slightly smaller than in the $\ell=2$ case at $U=3$, suggesting that the $P^{tot}_{d}$ in $\ell=3$ is expected to be attenuated compared to $\ell=2$ when $U=3$ and $V,J=0$. Looking at $a_{[0,0,2]}$ and $a_{[0,1,2]}$, although being the dominant coefficients at $U,V,J=1$, they are now reduced to secondary contributions. A contrasting scenario is observed where $a_{[0,0,2]}$ and $a_{[0,1,2]}$ are slightly larger in $\ell=3$ compared to the $\ell=2$ case. Along with other coefficients providing non-negligible contributions in the presence of finite $J$, the weakening of $a_{[2,0,0]}$ and $a_{[3,0,0]}$ coefficients in the $\ell=3$ system is counteracted, resulting in an equivalent $P^\parallel_{d}$ between $\ell=2$ and $\ell=3$ at $J=0.3$, as shown in Fig.~\ref{fig:para_coeff}(c).

\begin{figure}[h]
    \centering
    \includegraphics[width=1\linewidth]{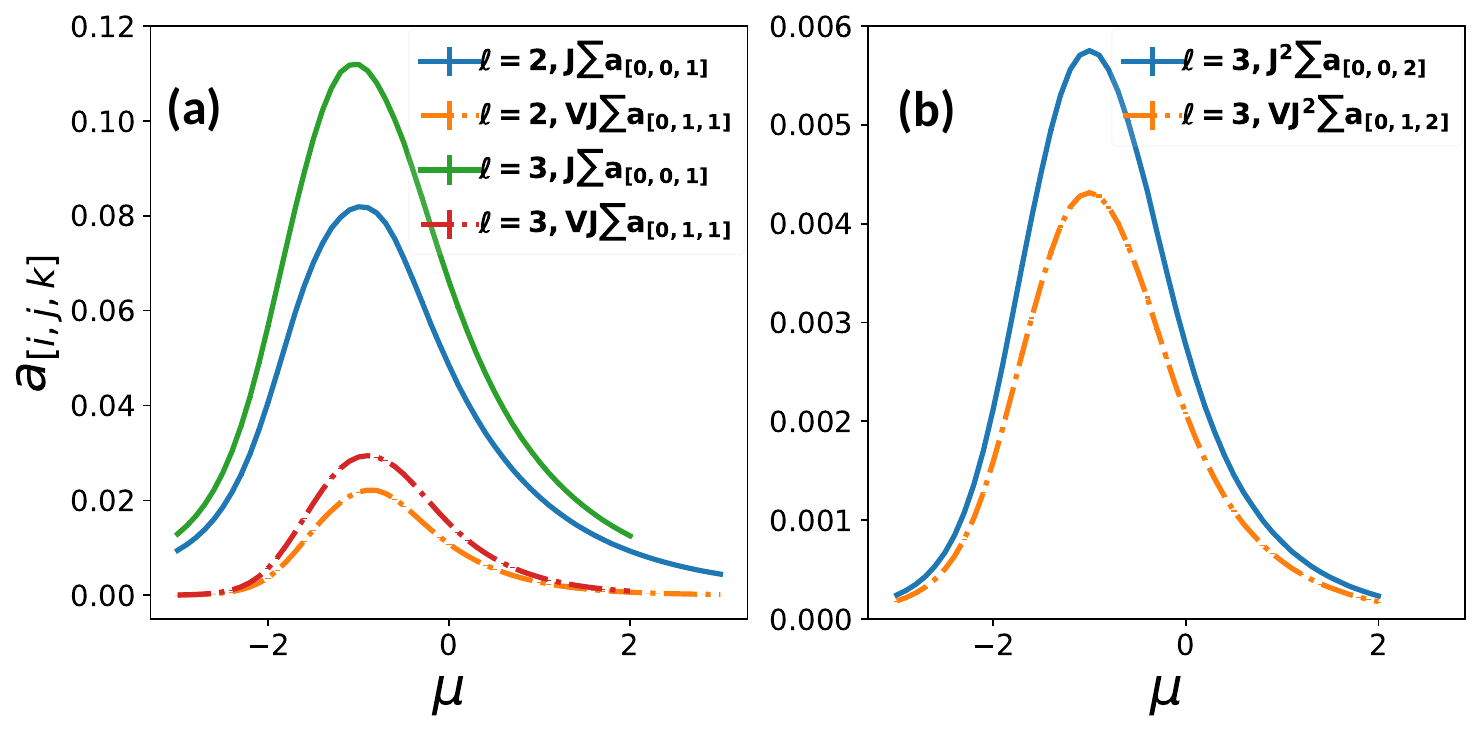}\\%
    \vspace{-0.14cm}\includegraphics[width=0.9\linewidth]{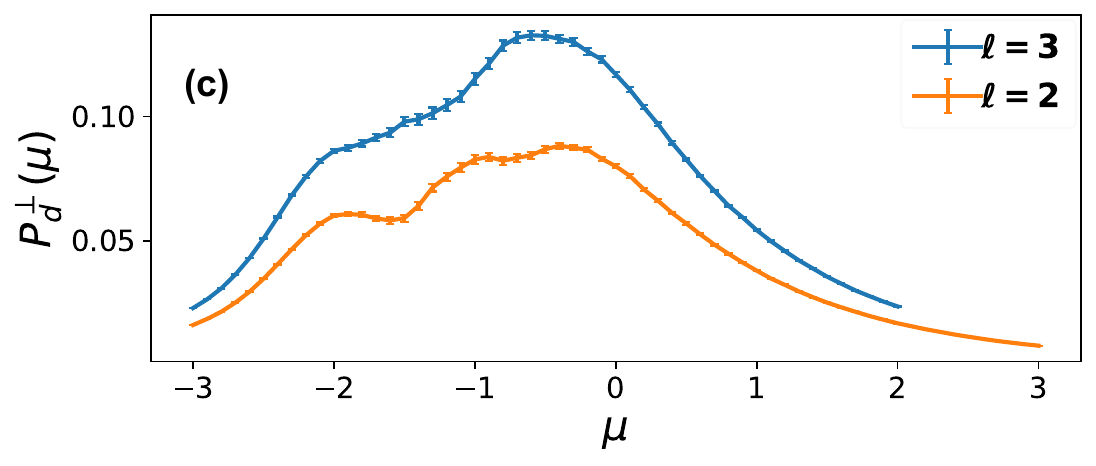}
    \caption{The overall contribution of the dominant coefficients in $\ell=2,3$ to $P^{\perp}_{d}$ for $U=3,V=0.75,J=0.3$  as function of $\mu$ for (a) $a_{[0,0,1]}$ with dashed line and $a_{[0,1,1]}$ with solid line (b) $a_{[0,1,2]}$  with dashed line and  $a_{[3,0,0]}$ with solid line. Note that the coefficients in (b) arising from the next-adjacent off-diagonal channels are absent in $\ell=2$. (c) Comparison of $P^{\perp}_{d}$ between $\ell=2$ and $\ell=3$ as a function of $\mu$ with the same choice of interaction strength.}
    \label{fig:perp_coeff}
\end{figure}

The adjacent-plane off-diagonal channels, $P^{0,\frac{\pi}{2}}_{d}$ and $P^{\frac{\pi}{2},\pi}_{d}$ in $\ell=3$, also retain the same leading coefficients consisting of identical diagram topologies as in the $P^{0,\pi}_{d}$ in $\ell=2$ case. To asses their impact on $P^{\perp}_{d}$, we now plot the overall contribution of  $a_{[0,0,1]}$ and $a_{[0,1,1]}$ coefficient  for  $U=3.0$, $V=0.75$ and $J=0.3$ in Fig.~\ref{fig:perp_coeff}(a). It is evident that these coefficients in $\ell=3$ contribute significantly more in  $P^{\perp}_{d}$ than in $\ell=2$. This increased contribution is attributed to the presence of two pairs of adjacent off-diagonal channels ($P^{0,\pi/2}_{d}$,$P^{\pi/2,0}_{d}$ \& $P^{\pi/2,\pi}_{d}$,$P^{\pi,\pi/2}_{d}$) containing these coefficients in $\ell=3$ as opposed to a single pair ($P^{0,\pi}_{d}$,$P^{\pi,0}_{d}$) in $\ell=2$. The coefficient sets in these channels have comparable pairing amplitude for both $\ell=2$ and $\ell=3$. Consequently, when the increased number of channels is summed and normalized by $1/\ell$, a net additive effect is yielded. 

The only difference between $\ell=2$ and $\ell=3$ lies in the presence of the next-adjacent off-diagonal $P^{0,\pi}_{d}$ channels, whose non-zero coefficients $a_{[0,0,2]}$ and $a_{[0,1,2]}$ have positive contributions, albeit weaker than dominant $a_{[0,0,1]}$ and $a_{[0,1,1]}$ coefficients.  Note that $a_{[0,0,2]}$ consists of a single second-order ladder diagram while $a_{[0,1,2]}$ has collections of multiple topologies. We plot the contribution of these coefficients from the next-adjacent off-diagonal channels to $P^{\perp}_{d}$ in Fig.~\ref{fig:perp_coeff}(b) and show that they further help enhance $P^{\perp}_{d}$ in $\ell=3$. Therefore, we arrive at a conclusion that in the presence of $J$, $P^{\perp}_{d}(\ell=3) > P^{\perp}_{d}(\ell=2)$ as demonstrated in Fig.~\ref{fig:perp_coeff}(c). Nevertheless, our comparative analysis of the coefficients indicates diagrammatic processes that enhance the pairing in the bilayer to be similar to those in the trilayer case.

\begin{figure}[ht]
    \centering
   \includegraphics[width=1\linewidth]{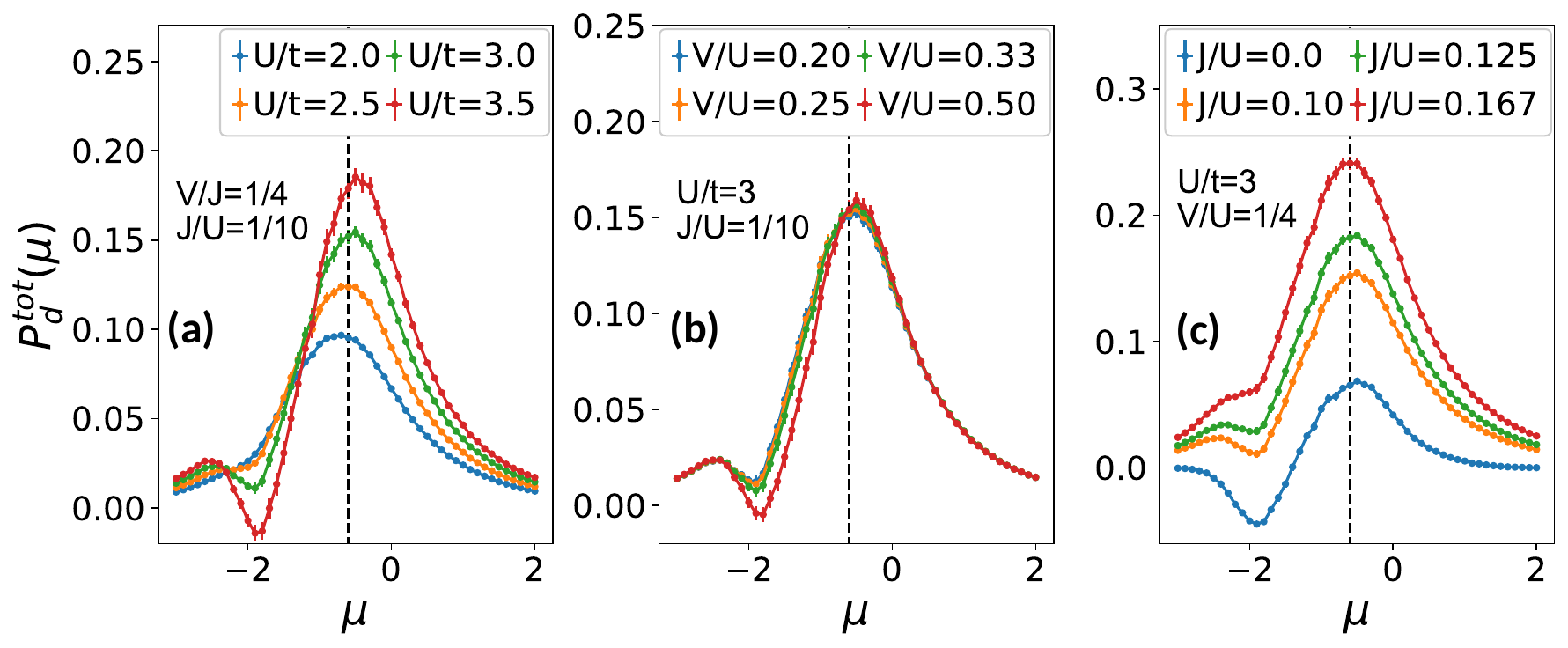}\\%
    \vspace{0.05cm}\includegraphics[width=1\linewidth]{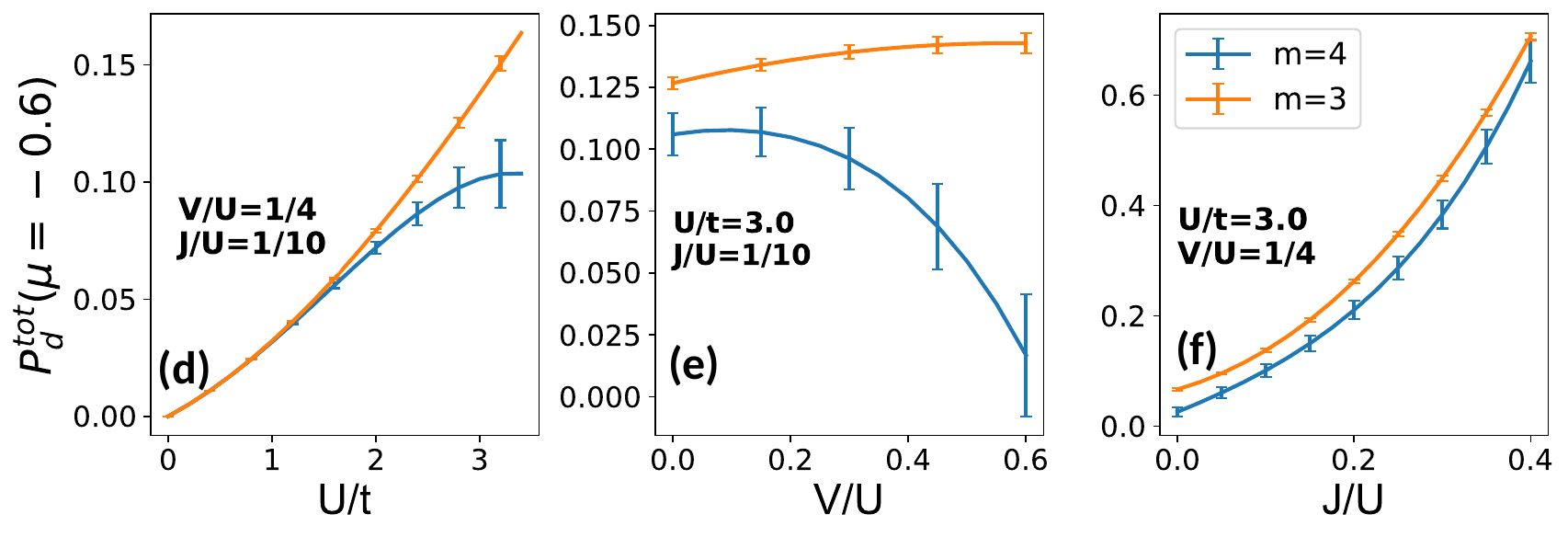}
    \caption{\label{UVJ_Ptot1} $P^{tot}_{d}$ for $\ell=3$ obtained by summing the $P^{\perp}_{d}$ and $P^{\parallel}_{d}$ components  is plotted as a function of $\mu$ (a-c). The blacked dotted line at $\mu=-0.6$ shows where truncated 3rd order $(m=3)$ $P^{tot}_{d}$ is peaked for $U/t=3$ case. Fourth-order correction $(m=4)$ is applied to $P^{tot}_{d}$ and plotted as a function of (d) $U/t$, (e)$V/U$ and (f) $J/U$ at $\mu=-0.6$.  } 
\end{figure}

For completeness, we plot $P^{tot}_{d}$ as a function $\mu$ in Fig.~\ref{UVJ_Ptot1}(a-c) with same parameter choice as the $\ell=2$ in Fig.~\ref{UVJ_Ptot}(a-c). Apart from exhibiting greater sensitivity to $J$, there are no particular notable distinctions. 
We also assess the effect of fourth-order correction on a single data point owing to the massive computational expense that is more severe in the trilayer case as indicated by Tab.~\ref{table1}. We apply fourth-order correction to   $\mu=-0.6$ in Fig.~\ref{UVJ_Ptot1}(d-f), where the peak is located to maintain consistency with the $\ell=2$ case.  Our results indicate only one notable distinction compared to the $\ell=2$ system.  With increasing $V$, the fourth correction exhibits a significant weakening of $P^{tot}_{d}(\ell=3)$ where it is fully suppressed for the large $V/U=0.5$ we utilized, in contrast to weakly dependent $V$ features in $P^{tot}_{d}(\ell=2)$. This suggests that overlooking the $V$ interaction as it can done for bilayer cases is not ideal, especially for large $V$. Of interesting note, for large $J/U=0.4$,  $\ell=3$ system shows an approximate seven-fold increase in $P^{tot}_{d}$ compared to a fourfold increase observed in the $\ell =2$ system.

Using the truncated fourth order $P^{tot}_{d}$ results on $\ell=2,3$ systems, we can establish a region of $U/t \leq 3$ and $V/U \leq 0.3$ on interaction space in which our third-order perturbative expansion remains controlled, and fourth-order corrections are minimum. While $J/U$ has a wider region of validity, we restrict ourselves to $J \leq 1/6$ so that it remains realistic and comparable to Hund's coupling.  Confining ourselves to this parameter space allows us to study layer dependency with third-order expansion in the proceeding section.

Finally, we conclude our discussion with remarks on the inter-layer pairing $d^{(1)}_{z}$ and $d^{(2)}_{z}$ in $\ell=2$ and $\ell=3$ system. In the context of trilayer cuprates, inter-layer pairing alongside Cooper pair hopping is considered a possible mechanism behind enhancement in $T_{c}$  observed experimentally \cite{trilayer_cuprate,trilayer_cuprate(1)}. In our calculation (not shown), however, we have found inter-layer pairing ($d_{z}$) to be entirely repulsive in the truncated third order in the weak coupling \emph{U-V-J} space at  $t_{\perp}=0.125$ and $\beta=5$. Although a positive fourth-order correction leads to an attractive region in the intermediate coupling range, access to a higher-order expansion is necessary to attain reliable results, which is beyond the scope of the study.

\subsection{ Quadlayer ($\ell=4$) result and beyond} 

Having analyzed the $\ell=2$ and $\ell=3$ pairing, extending beyond allows us to generate a more comprehensive picture of the layer-dependent features in $P^{tot}_{d}$. To do so, we first extend our study by stacking one additional layer to the $\ell=3$ system, resulting in the so-called quad-layer model ($\ell=4$). The $P^{\parallel}_{d}$ component in $\ell=4$  system consists of four diagonal channels arising from the four momenta planes $P^{0,0}_{d}$, $P^{\frac{\pi}{3},\frac{\pi}{3}}_{d}$, $P^{\frac{2\pi}{3},\frac{2\pi}{3}}_{d}$ and $P^{\pi,\pi}_{d}$. The  $P^{\perp}_{d}$  comprises of three inequivalent adjacent off-diagonal channels $P^{0,\frac{\pi}{3}}_{d}$,$P^{\frac{\pi}{3},\frac{2\pi}{3}}_{d}$ and $P^{\frac{2\pi}{3},\pi}_{d}$  and two inequivalent next adjacent $P^{0,\frac{2\pi}{3}}_{d}$  and $P^{\frac{\pi}{3},\pi}_{d}$ channels. There also exists $P^{0,\pi}_{d}$ next-next adjacent channel where the only possible coefficient is $a_{[0,0,3]}$ within the truncated third order. We found $a_{[0,0,3]}$  to have a negligible contribution, especially when expanded in powers of $J^3$ with $J<1$. So it can safely be ignored. Summing all the channels results in 

\begin{equation}
\begin{split}
P^{tot}_{d}(l=4) &= ( P^{0,0}_{d} + P^{\frac{\pi}{3},\frac{\pi}{3}}_{d} + P^{2\frac{\pi}{3},2\frac{\pi}{3}}_{d} + P^{\pi,\pi}_{d} ) / 4 \\
&\quad + ( P^{0,\frac{\pi}{3}}_{d} + P^{\frac{\pi}{3},\frac{2\pi}{3}}_{d} + P^{2\frac{\pi}{3},\pi}_{d} ) / 2 \\
&\quad + ( P^{0,\frac{2\pi}{3}}_{d} + P^{\frac{\pi}{3},\pi}_{d} ) / 2.
\end{split}
\end{equation}
here the first term represent $P^{\parallel}_{d}$ and the last two terms represents $P^{\perp}_{d}$.

\begin{figure}
    \centering
    \includegraphics[width=1\linewidth]{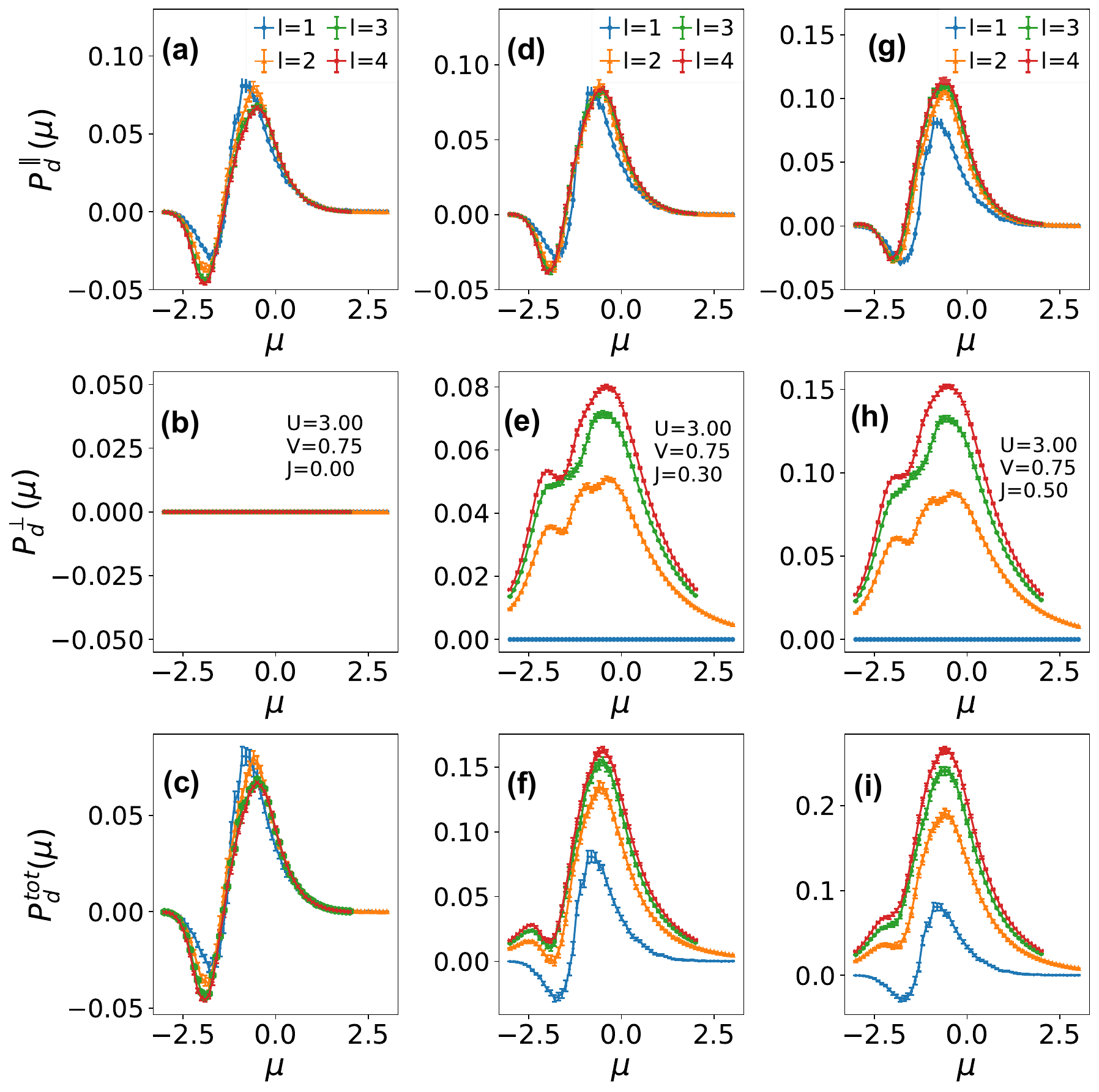}
    \caption{\label{layer_dep}  $P^{\parallel}_d$, $P^{\perp}_d$ and $P^{tot}_d$ is plotted  for up to $\ell =4$ as a function of $\mu$. With fixed $(U,V)=(3.0,0.75)$, three choices of $J$ interaction is utilized  $J=0.0$ (a-c),$J=0.30$ (d-f) and $J=0.50$ (g-i).}
 \end{figure}

To quantify the layer dependence, we plot the $P^{\parallel}_{d}$, $P^{\perp}_{d}$ and $P^{tot}_{d}$  up to $\ell=4$ system as a function of $\mu$ with a fixed $U=3.0$ and $V=0.75$ for three choices of $J=0.0$, $J=0.3$ and  $J=0.5$ depicted in the three panels of Fig.~\ref{layer_dep}. In the absence of $J=0$ interaction, we see that $P^{tot}_{d}$ consisting only of $P^{\parallel}_{d}$ component is only slightly attenuated with the addition of layers Fig.~\ref{layer_dep}(a-c). This suggests that without $J$ hopping interaction, the pairing process in the layered model is qualitatively similar to those found in the single-layer Hubbard model. What sets the layered model apart is the presence of $J$ interaction, particularly 
attractive nonlocal pair hopping $J'$, as evident in Fig.~\ref{layer_dep}(f,i). When a small $J=0.3$ is utilized, enhancement received by $P^{\parallel}_{d}$ offsets the weak attenuation observed in $J=0$ and magnitude of $P^{\parallel}_{d}$ becomes comparable among the four cases we study. For a larger value of $J=0.5$, we observe that the $P^{\parallel}_{d}$ for $\ell>1$ is larger and separates itself from the $\ell=1$ case. However, among $\ell=2,3,4$, $P^{\parallel}_{d}$  remains roughly the same as in the case of $J=0.3$. Thus, we conclude that $P^{\parallel}_{d}$ plays no role in the layer dependency in the $\ell \geq 2$ system. On the contrary,  $P^{\perp}_{d}$ sees a substantial enhancement with increasing layers that is dependent on the strength of  $J$ used as portrayed in Fig.~\ref{layer_dep}(e,h). However, this growth is sublinear, indicated by the saturating tendency with $\ell$.

  \begin{figure}
    \centering
   \includegraphics[width=0.95\linewidth]{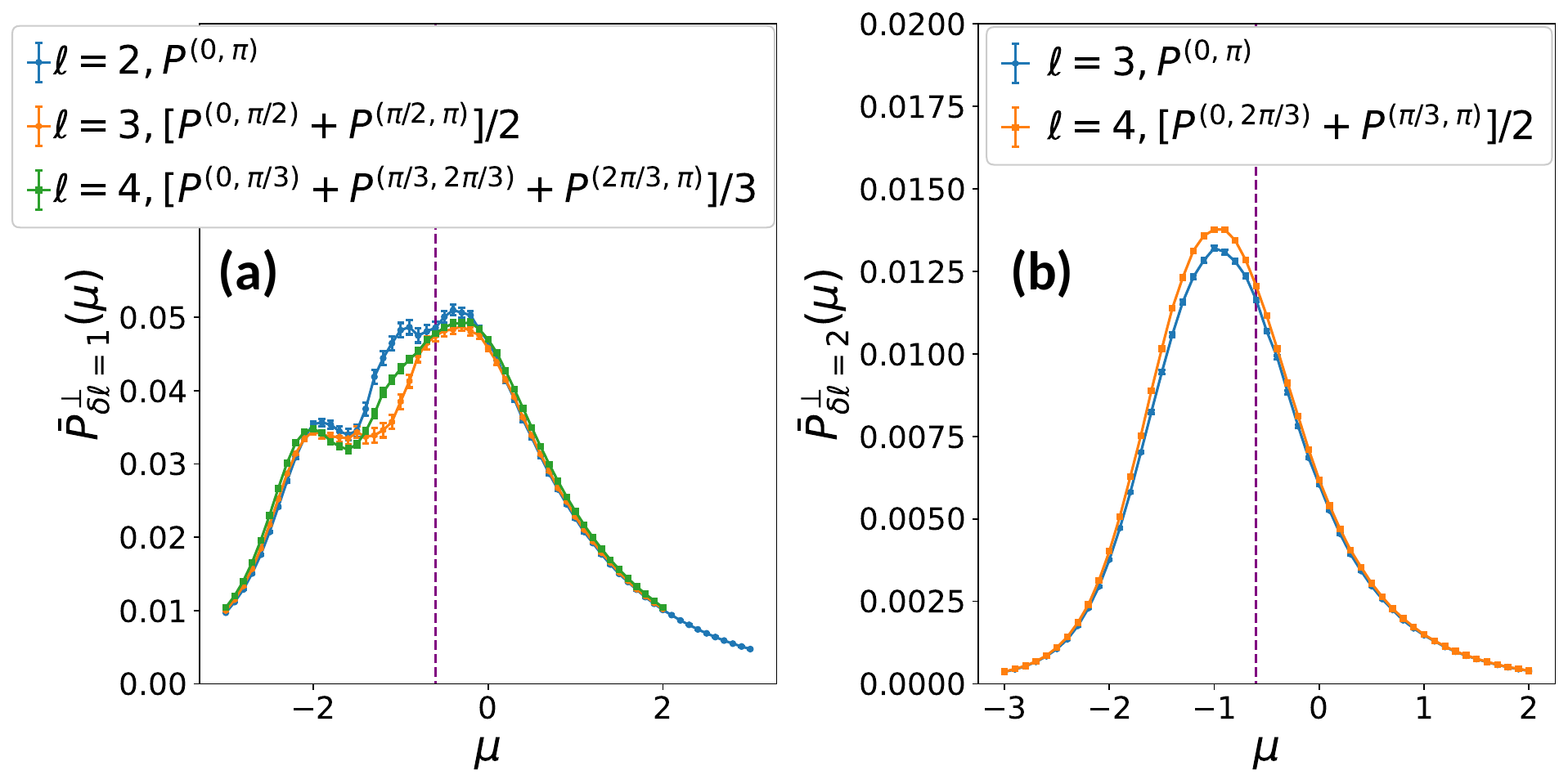}\\%
    \includegraphics[width=0.98\linewidth]{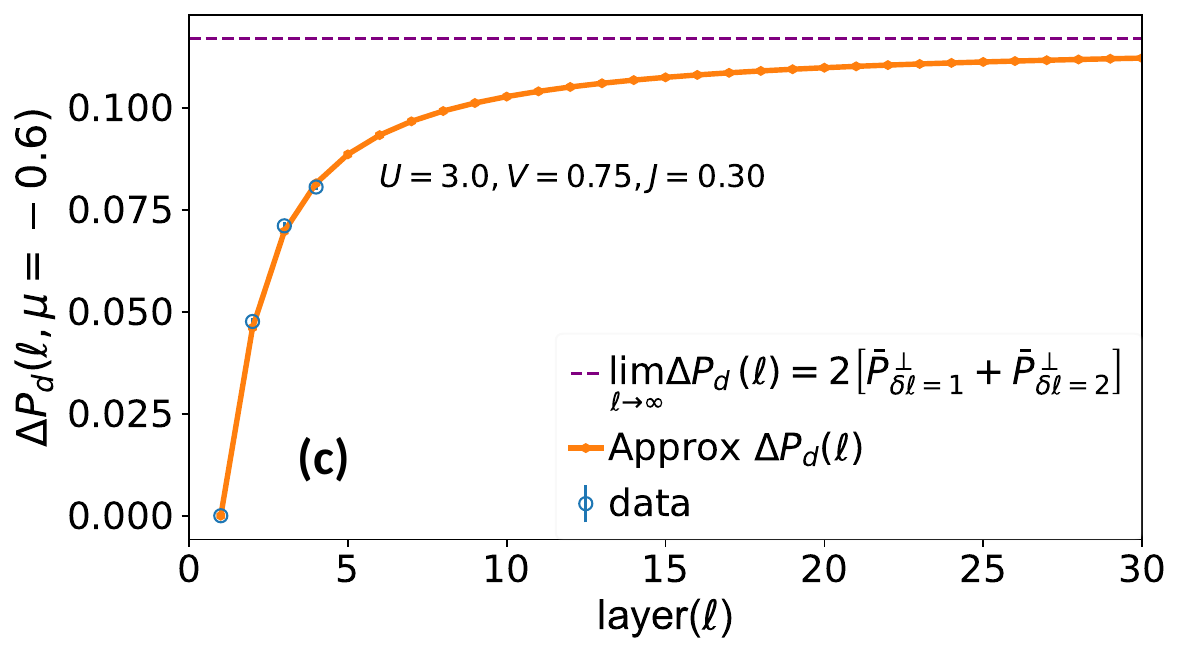}
    \caption{\label{fig:average} (a,b)  averaged adjacent inter-plane scattering  term $\bar{P}^{\perp}_{\delta \ell =1}$ and averaged next-adjacent inter-plane scattering term $\bar{P}^{\perp}_{\delta \ell =2}$ plotted as a function of $\mu$ for up to $\ell=4$. The vertical dotted line represents the point $\mu=-0.6$ where these two terms are roughly constant among the $\ell$. (c) The solid orange line plots the approximate $\Delta{P}_{d}(\ell)$ from the generalized layer-dependent equation in Eq.~\ref{eq:ldep} arising from $P^{\perp}_{d}$ component due to c-axis pair hopping. The horizontal purple dashed line indicates the $\ell \to \infty$ limit. The unfilled blue circles depict the actual changes in pairing for $\ell = 2, 3, 4$.  } 
\end{figure}

To speculate the layer dependency of $P^{tot}_{d}$ beyond $l>4$ for finite $J$, we first make an assumption that there is no layer dependency $P^{\parallel}_{d}(\ell) \approx \textit{const}$  and that layer dependent features $P^{tot}_{d}$ are determined by only the  $P^{\perp}_{d} (\ell)\approx (P_{d}^{tot}(\ell)-P_{d}^\parallel )$ component.  This implies that Cooper pair hopping between the planes via $J$ and $J'$ interactions are the only physical processes within our model that give rise to layer-resolved features. This enables us to approximate $P^{\perp}_{d} \approx \Delta P_{d}(\ell)$, where $\Delta P_{d}(\ell)$ is a change in pairing as a function of $\ell$. Our prior analysis of the coefficient in $\ell=3$  showed that different $ P^{k_{z},k'_{z}}_{d}$ scattering processes decay in amplitude drastically based on their  $k_z$ separation in momenta planes. Scattering to the adjacent plane is dominant; the next adjacent plane is subdominant, and beyond, the contributions are negligible for small $J\leq0.5$. However, not all the same $k_{z}$ separation in $ P^{k_{z},k'_{z}}_{d}$ channels have the equivalent magnitudes and peak locations.  If the \emph{averaged} $P^{k_{z},k'_{z}}_{d}$ channels with the same $k_{z}$ separation are constant as a function of $\ell$ for a fixed $\mu$, a generalized equation that estimates $\Delta P_{d}(\ell)$ in the large $\ell$ limit can be formulated. This involves multiplying the averaged adjacent and next-adjacent channels by their respective numbers, scaling as $2(\ell-1)$ and $2(\ell-2)$, and normalizing by $1/\ell$. To do this, we introduce variables $\bar{P}^{\perp}_{\delta \ell =1}$ and $\bar{P}^{\perp}_{\delta \ell =2}$, representing the average of adjacent channels ($\delta \ell =1$) and next-adjacent channels ($\delta \ell =2$) with the same $k_{z}$ separation. For example, in $\ell=4$, $\bar{P}^{\perp}_{\delta \ell =1}$ is the average of three adjacent channels: $\bar{P}^{\perp}_{\delta \ell =1} = (P^{0,\pi/3}_{d} + P^{\pi/3,2\pi/3}_{d} + P^{2\pi/3,\pi}_{d})/3$. Similarly, $\bar{P}^{\perp}_{\delta \ell =2}$ is the average of two next-adjacent channels: $\bar{P}^{\perp}_{\delta \ell =2} = (P^{0,2\pi/3}_{d} + P^{\pi/3,\pi}_{d})/2$.  In Fig.~\ref{fig:average}(a,b), we plot $\bar{P}^{\perp}_{\delta \ell =1}$ and  $\bar{P}^{\perp}_{\delta \ell =2}$ for $\ell=2,3,4$ systems and find existence of wide range of $\mu$ where the curves overlap within the errorbar, indicating that two variables can be approximated as a constant. Generalizing this to $\ell$ layers and  treating  $\bar{P}^{\perp}_{\delta \ell =1}$ and $\bar{P}^{\perp}_{\delta \ell =2}$ as a constant, one can formulate an equation as a combinatorics problem that estimates the change in pairing with  as a function of the layers with respect to the single layer model, given by

\begin{align}
    \Delta P_{d}(\ell) &= \frac{2(\ell-1)}{\ell}\bar{P}^{\perp}_{\delta \ell =1} + \frac{2(\ell-2)}{\ell}\bar{P}^{\perp}_{\delta \ell =2} \nonumber \\
    &= 2\left[\left(1 - \frac{1}{\ell}\right) \bar{P}^{\perp}_{\delta \ell =1} + \left(1 - \frac{2}{\ell}\right)\bar{P}^{\perp}_{\delta \ell =2}\right].
    \label{eq:ldep}
\end{align}

  Here we extract the values of $\bar{P}^{\perp}_{\delta \ell =1}$ and  $\bar{P}^{\perp}_{\delta \ell =2}$ from $\mu=-0.6$ where the both variables are roughly equal across the $\ell$ layers as indicated by the dotted line in Fig.~\ref{fig:average}(a,b) and  plot Eq.\ref{eq:ldep} as a function of $\ell$ in in Fig.~\ref{fig:average}(c)  at $U=3$, $V=0.75$ and $J=0.3$. As a reference to readers, we also plot results for $P^{\perp}_{d}$ for $\ell=1,2,3,4$. We emphasize here that this equation should not be treated as a fitting to the data. But rather to capture how paring is expected to change as a function of $\ell>4$ due to c-axis pair hopping. Our model indicates that  $\Delta P_{d}$ grows as a function of the layer sublinearly before saturating at the 3D limit where $\lim_{\ell \to \infty} \Delta P_{d}(\ell) = 2 \left[\bar{P}^{\perp}_{\delta \ell =1} + \bar{P}^{\perp}_{\delta \ell =2}\right]$. Therefore, one cannot attain a continuous phase transition at a fixed temperature just by stacking layers to the 3D limit despite the additive effect of off-diagonal pairing channels for the parameter space we explored. Nevertheless, the Eq.\ref{eq:ldep}  is reminiscent of the $ \Delta T_c(\ell) = \textit{const}\times(1 - 1/\ell)$ relation, derived by Legget from interplane Coulombic interaction in identically doped c-axis layered structures \cite{legget(TC)}. Nishiguchi  \emph{et al.} also reported a sublinear increase in $T_{c}$ in the layered Hubbard model up to $\ell=3$  using fluctuation exchange approximation (FLEX)  system by considering pair hopping of various configurations \cite{pairhopping(3)}. Chakravarty derived a similar relation via the inter-layer tunneling mechanism \cite{Chakravarty1998}.

While the results of our perturbative calculations are not directly applicable to superconducting cuprates at strong coupling and low-temperature limits, it is worth noting that the monotonic increase in $\Delta P_{d}$ with $\ell$ observed here contrasts with experimental findings for cuprates in the same homologous series. In cuprates, $T_{c}$ increases with the number of $CuO$ layers in the unit cell up to $\ell=3$, then decreases and eventually saturates at a certain value \cite{Tc_n(1), Tc_n(2)}. It is known that the electron and hole doping levels vary significantly between inner and outer $CuO$ layers in real and reciprocal space, even in the high-temperature phases \cite{mukuda,exp(hole_electro),trilayer_cuprate}. Within our model the  inter-layer tunneling induces layer-differentiated doping even in the absence of interactions at fixed $\mu$. At $\mu = -0.6$, where pairing is most enhanced, the corresponding bare densities $\vec{n}_{k_{z}}$ in the momentum layers are $n_{0} = 1.02$ for $\ell = 1$, $(n_{0}, n_{\pi}) = (1.10, 0.92)$ for $\ell = 2$, $(n_{0}, n_{\frac{\pi}{2}}, n_{\pi}) = (1.12, 1.02, 0.86)$ for $\ell = 3$, and $(n_{0}, n_{\frac{\pi}{3}}, n_{\frac{2\pi}{3}}, n_{\pi}) = (1.13, 1.07, 0.97, 0.84)$ for $\ell = 4$ systems, respectively. We observe a general trend where outer momenta layers prefer to be oppositely doped in the bare electronic structure. We have also calculated the renormalized $n_{k_{z}}$ and momentum distribution of filling using the second-order self-energy for $\ell=2, 3,$ and $4$. The renormalization with a finite $ U=3$ causes a small shift in density, but it remains comparable to the bare case and is largely insensitive to the $V$ and $J$ interactions. The results for both the bare and renormalized densities are discussed in the supplemental materials and we see that bare densities remain a good estimate in the weak coupling regime. The average band filling, $n_{\text{avg}} $, for both the bare and renormalized cases in layers $\ell = 1, 2, 3$, and $4$ is close to half-filling, with $ n_{\text{avg}} $ values ranging betwen 0.96 to 1.01.

 We have also studied the effect of varying density in the reciprocal layers by changing the $\mu$ independently in the bands for the $\ell=2$ and $\ell=3$ system in supplemental. We assess whether the fermi surface can be tuned to take advantage of the van Hove singularity (VHS) to promote pairing \cite{VHS:chen,farid:2023}. We have identified distinct $\vec{\mu}$ regions where pairing is either attractive or repulsive, with varying amplitudes. The pairing response is strongest when $\mu$ is located in the proximity the VHS of respective bands. We determine where the optimal $\vec{\mu}$ occurs based on where $P^{tot}_{d}$ is peaked and report the corresponding $n_{k_{z}}$ densities. However, fine-tuning $\vec{\mu}$ to VHS does not lead to an appreciable rise in $P^{tot}_{d}$ for $\ell=2$ and $\ell=3$ system at $J=0$, with $J$ playing a significantly more prominent role. 

It is evident that `out-of-plane' c-axis pair hopping when induces a clear layer-dependent features in $P^{\perp}_{d}$ within the weak-coupling $U$-$V$-$J$ interaction and $\mu$ space we explore. The choice of isotropic versus anisotropic hybridization has minimal impact on this behavior. Further, the layer-dependent feature in $P^{\perp}_{d}$ primarily originates from the dominant $a_{[0,0,1]}$ and $a_{[0,0,2]}$ coefficients in adjacent and next-adjacent off-diagonal channels in $P^{\perp}_{d}$, as shown in Fig.~\ref{fig:perp_coeff} for $\ell=2$ and $\ell=3$ systems. These coefficients form a ladder topology that is independent of $U$ interaction.  Therefore,  even in the absence of $U$, a monotonic increase in pairing akin to Fig.~\ref{fig:average}(c)  can be obtained simply via layer stacking when $J'/J <0$. Again, when $U$ is finite and small, a layer-dependent feature is always expected to appear in $P^{\perp}_{d}$ component even in the asymptotic $J \to 0$ limit. Consequently, our results are robust and applicable to weakly interacting generic layered structures as long as c-axis pair hopping is allowed and $\mu$ in the reciprocal layer is fixed.

\subsection{Summary and Outlook }
In this work, we have introduced a new symbolic technique to iteratively assign the band indexes to topologies to generate complete symbolic representations of diagram sets responsible for $P^{k_{z}, {k'_{z}}}_{d}$. We resolved Matsubara frequency summations analytically using algorithmic Matsubara integration and momentum dependency stochastically. For each $P^{k_{z}, {k'_{z}}}_{d}$, we categorized the diagram sets based on interaction to obtain a set of coefficients in the multi-power-series expansion giving full access to weak-coupling $U,V,J$ space. We have summed the diagonal and off-diagonal channels separately to obtain two independent components of pairing, $P^{\parallel}_{d}$  and  $P^{\perp}_{d}$  with total physical pairing susceptibility being $P^{tot}_{d} = P^{\parallel}_{d}+P^{\perp}_{d}$. We have probed $P^{\parallel}_{d}$, $P^{\perp}_{d}$ and $P^{tot}_{d}$ as a function of $\mu$ for up to an $\ell=4$ system in the presence of local $U,V,J$ interactions with a nonlocal pair hopping $J'$ fixed at  $J'/J=-0.5$. 

From the analysis of coefficients in $\ell=2$ and $\ell=3$ diagonal channels, we have determined that coefficients $a_{[2,0,0]}$ and $a_{[3,0,0]}$ consisting of only Hubbard  interaction are the dominant contributor to $P^{\parallel}_{d}$ at $U=3$. The coefficients $a_{[0,0,2]}$ and $a_{[0,1,2]}$ facilitate the enhancement in $P^{\parallel}_{d}$ component that scaling quadratically with $J$. When $J=0$, we showed that $P^{\parallel}_{d}$ is weakly attenuated with $\ell$ but qualitative features remain similar to $\ell=1$ system. For a finite $J$, the enhancement received across the layer is the same and exhibits weak to no layer-resolved features beyond $\ell=1$ for $P^{\parallel}_{d}$. This is attributed to the number of diagonal channels being fixed at $\ell$, where the contribution from each channel is averaged out by the normalization factor of $1/\ell$.

The contribution to $P^{\perp}_{d}$  in small $J/U\leq 1/6 $ that is comparable to Hund's coupling predominately comes from dominant $a_{[0,0,1]}$ and $a_{[0,0,2]}$ coefficients in adjacent and next adjacent off-diagonal channels consisting of the first order and second order ladder diagram respectively. Scattering at and beyond $J^3$ is negligible, and we see that an attractive non-local interaction $J'/J<0$ is a crucial requirement for the coefficients to be positive and dominant. The channels possessing these coefficients grow super-linearly with $\ell$, yielding a net additive effect even when normalized by $1/\ell$, and are thus responsible for layer-dependent features in pairing.  We find that even a small $J/U=0.10$ results in a nearly two-fold increase in $P^{tot}_{d}$ in the $\ell=3$ system compared to $\ell=1$, primarily arising from the $P^{\perp}_{d}$ component. We find this increment is monotonic but sublinear as a function of layer till $\ell=4$. To speculate beyond $\ell=4$, we assume that $ P^{\parallel}_{d}(l) \approx \textit{const}$  and use the observation that the averaged adjacent $\bar{P}^{\perp}_{\delta \ell =1}$ and next adjacent $\bar{P}^{\perp}_{\delta \ell =2}$  off-diagonal channels in $P^{\perp}_{d}$ remain constant as a function of $\ell$ at  $\mu=-0.6$, where the pairing is largest. Knowing that the number of adjacent and next adjacent channels scales as  $2(\ell-1)$ and $2(\ell-2)$ as a function of $\ell$, one can formulate a generalized equation for $\Delta P_{d}(\ell)$ that predicts pairing trends beyond $\ell=4$. Our model indicates a well-defined 3D limit where pairing saturates. 

In the supplemental section, we have  calculated $q=(\pi,\pi)$ spin susceptibility for $\ell=2$ and found little to no dependency in $V, J$, and $J'$ interaction space. From this, it is suggested that spin fluctuations do not mediate the layer-resolved features in $P^{\perp}_{d}$. We  show that bare electronic structure of our model has layer differentiated doping that is induced by inter-layer hybridization $t_{z}$. We further vary the doping by changing the $\mu$ distinctly in the momenta layers and find that the optimal region lies near the van Hove singularity.  A detailed discussion on optimal $\mu$ and the corresponding bare and renormalized densities are provided in the supplemental. However, adjusting solely for optimal $\mu$ does not lead to appreciable rise in $P^{tot}_{d}$ in comparison to the influence of pair hopping.

Our calculations unequivocally indicate that both local and non-local pair hopping plays a significant role in promoting pairing monotonically as a function of layers when $J'\leq 0$ and suppressing pairing when $J'> 0$  in a generic weakly interacting layered structure at high-temperature. Finally, we emphasize that our approach has potential applications in understanding pairing processes in any material that has a low-energy multi-band effective model with any arbitrary long-range interaction. In particular, our methodologies can be readily extended to study single and two-particle properties in more realistic systems like the single layer three-band Emery model or Sr$_2$RuO$_4$, enabling us to identify key diagrammatic processes and coefficients within weak coupling regimes \cite{SrO1, SrO2,emery1,wmery2}.

\section{Acknowledgement}
We acknowledge the support of the Natural Sciences and Engineering Research Council of Canada (NSERC) RGPIN-2022-03882. Our codes make use of the ALPSCore\cite{ALPSCore,alpscore_v2} and AMI libraries\cite{libami,torchami}.

\renewcommand{\thefigure}{S\arabic{figure}} 
\renewcommand{\thetable}{S\arabic{table}}  
\setcounter{figure}{0}  
\setcounter{table}{0}
\section*{Supplementary Information}
\subsection{Non-interacting density of states and filling in the reciprocal layers}

\begin{figure}[ht]
  \centering
  \includegraphics[width=1\linewidth]{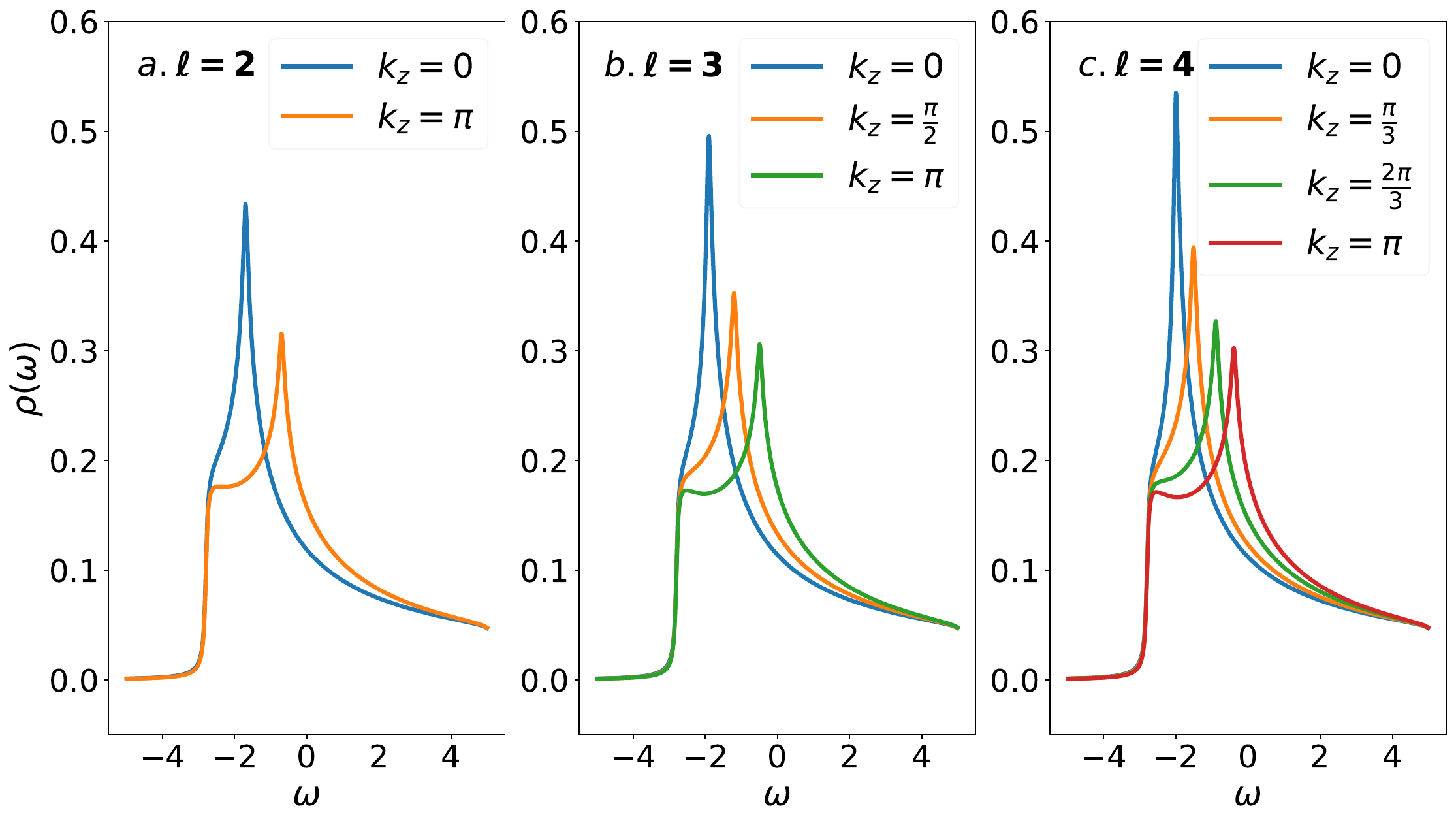}
  \caption{Non-interacting density of states $\rho(\omega)$  for (a) $\ell=2$, (b) $\ell=3$, (c) $\ell=4$ at $(t_{\perp},t_{bs},t^\prime,\mu) = (0.125,0.0,-0.3,0.0)$ and $U,V,J=0$. The van Hove singularities are located at chemical potentials $\vec{\mu}=-(0.7, 1.7)$,  $\vec{\mu}=-(0.5, 1.2, 1.9)$, and $\vec{\mu}=-(0.4,0.9,1.5,2.0)$ for $\ell=2$, $\ell=3$ and  $\ell=4$, respectively }
  \label{fig:fermi}
\end{figure}

\begin{figure}[ht]
  \centering
  \includegraphics[width=1\linewidth]{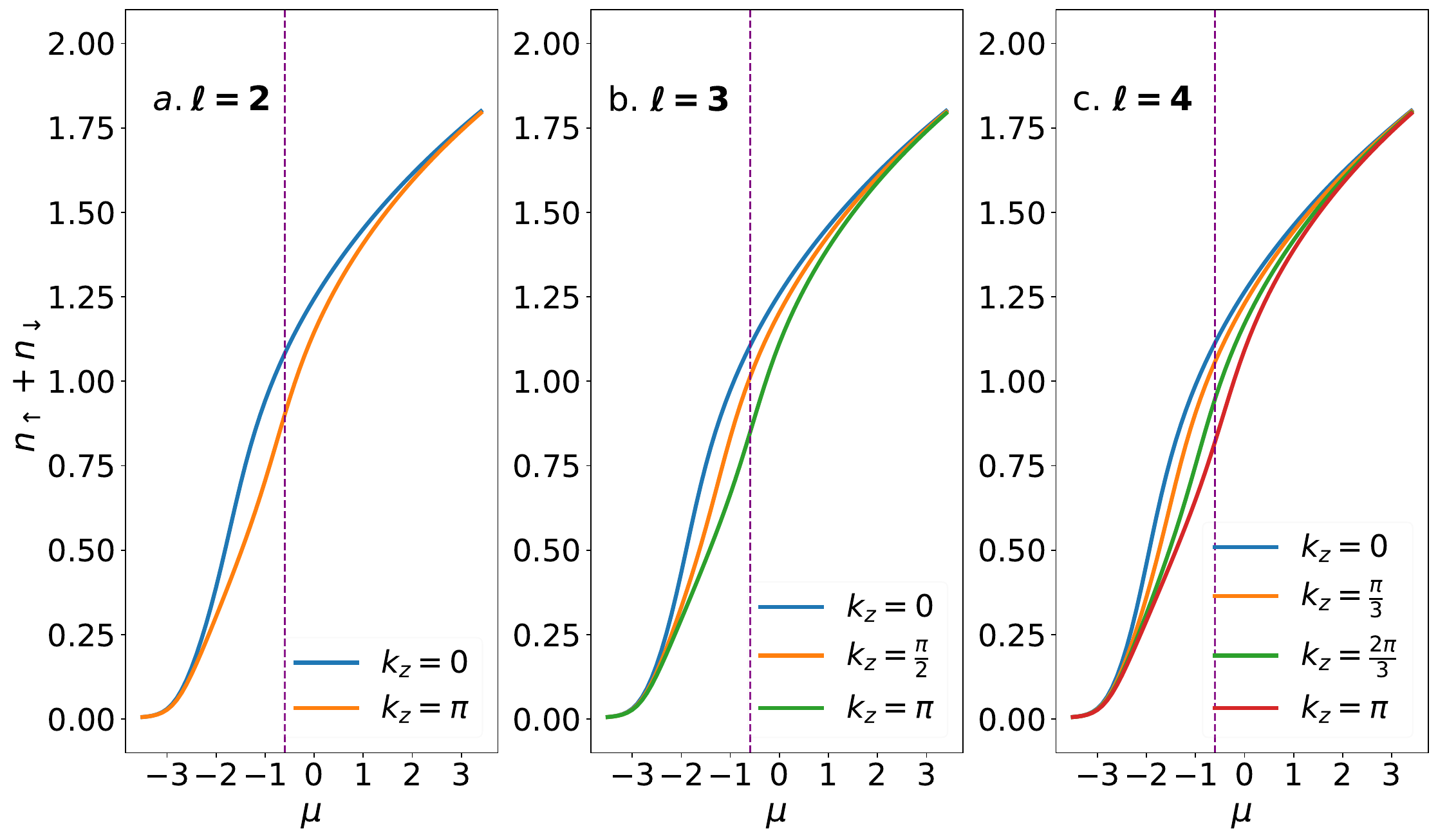}
  \caption{Spin summed density $n$ as a function of  layer $k_{z}$ and chemical potential $\mu$ for $\ell=2$, $\ell=3$ and $\ell=4$ at $\beta =5$ and $(t_{\perp},t_{bs},t^\prime) = (0.125,0.0,-0.3)$. The purple dotted line indicates $n$ at $\mu=-0.6$  }
  \label{fig:density}
\end{figure}

The density of states $\rho(\omega)$ in the  momenta layers $k_{z}$ is given by

\begin{equation}
\rho(\omega) = -\frac{1}{N\pi} \sum_{\mathbf{q}} \text{Im}\left[ G_{k_z}(\mathbf{q}, \omega + i\eta) \right],    
\end{equation}
where $N = L \times L$ is the size of the lattice, $\mathbf{q} = (q_x, q_y)$ is the external momenta, and $\eta$ is an infinitesimally small positive number required for analytical continuation ($\eta \to 0^+$). Using $L = 86$ and $\eta = 0.01$, we plot non-interacting  $\rho(\omega)$ for $\ell = 2,3$ and $4$ systems in the three panels of Fig.~\ref{fig:fermi}. We utilize the same tight-binding parameters as in main text with $t_{\perp} = 0.125$, $t_{bs} = 0$ and $t^\prime = -0.3$. We observe that $\rho(\omega)$ features differ among the $k_{z}$ bands and layers, each possessing distinct van Hove singularities. However, when $t_{z} = 0$, the momenta layers are completely decoupled and will have the same $\rho(\omega)$. With the introduction of a finite $t_{z}$, the features of $\rho(\omega)$ in the different momenta layers begin to shift, and this deviation is determined by the strength of $t_{z}$. This effect creates the potential for layer-differentiated doping even in the absence of interactions. 
To quantify this, we calculate the spin summed density $n_{k_{z}} = n_{k_{z}\uparrow} + n_{k_{z}\downarrow}$ as a function of $\mu$ in Fig.~\ref{fig:density}, using the same parameters as in Fig.~\ref{fig:fermi}. The spin summed momentum distribution and density in a given $k_{z}$ momenta layer is calculated via
\begin{align}
n_{k_z}(\mathbf{q}) &= \frac{2}{\beta} \sum_{n\sigma} G_{k_{z}}(\mathbf{q}, i\omega_{n}) e^{i\omega_{n}0^{+}}\\
n_{k_z} &=   \frac{1}{N}\sum_{\mathbf{q}} n_{k_z}(\mathbf{q}).
\end{align}

From Fig.~\ref{fig:density}, we see that the presence of $t_{z}$ indeed results in layer-differentiated doping even in the absence of $U$, $V$, or $J$ interactions. We focus our discussion on $\mu=-0.6$ where the largest enhancement to pairing is attained as shown in the main text. At $\mu=-0.6$, corresponding bare densities are $(n_{0}, n_{\pi}) =(1.10, 0.92)$, $(n_{0}, n_{\frac{\pi}{2}}, n_{\pi}) = (1.12, 1.02, 0.86)$  and   $(n_{0}, n_{\frac{\pi}{3}}, n_{\frac{2\pi}{3}}, n_{\pi}) = (1.13, 1.07, 0.97, 0.84)$ for $\ell=2,3$ and $4$, respectively. We observe a general trend where the two outer most planes prefer to be oppositely doped. The $k_{z}=0$ momenta layers tend to be electron doped while $k_{z}=\pi$ hole doped. This change in doping occurs gradually as we move along $k_{z}$ separation, $\pi/(\ell-1)$. We note that for a fixed $\mu=-0.6$ in bands, the average band filling is close to half filling.

\subsection{ Renormalized momentum distribution and density in the reciprocal layer}
In the main text, the analysis of pairing susceptibility is primarily based on the truncated third-order pairing, $P_{d}^{\text{tot}}$. Within these particle-particle diagrams, there is a subset that includes the second-order self-energy insertion, $\Sigma^{(2)}$. This renormalization of the single-particle propagator is expected to modify the density, which is assessed in this section. Furthermore, we consider this modification as an estimate of the renormalized $n_{k_{z}}$ in the weak-coupling limit and compare it with the bare $n_{k_{z}}$.

\begin{figure}[ht]
  \centering
  \includegraphics[width=1\linewidth]{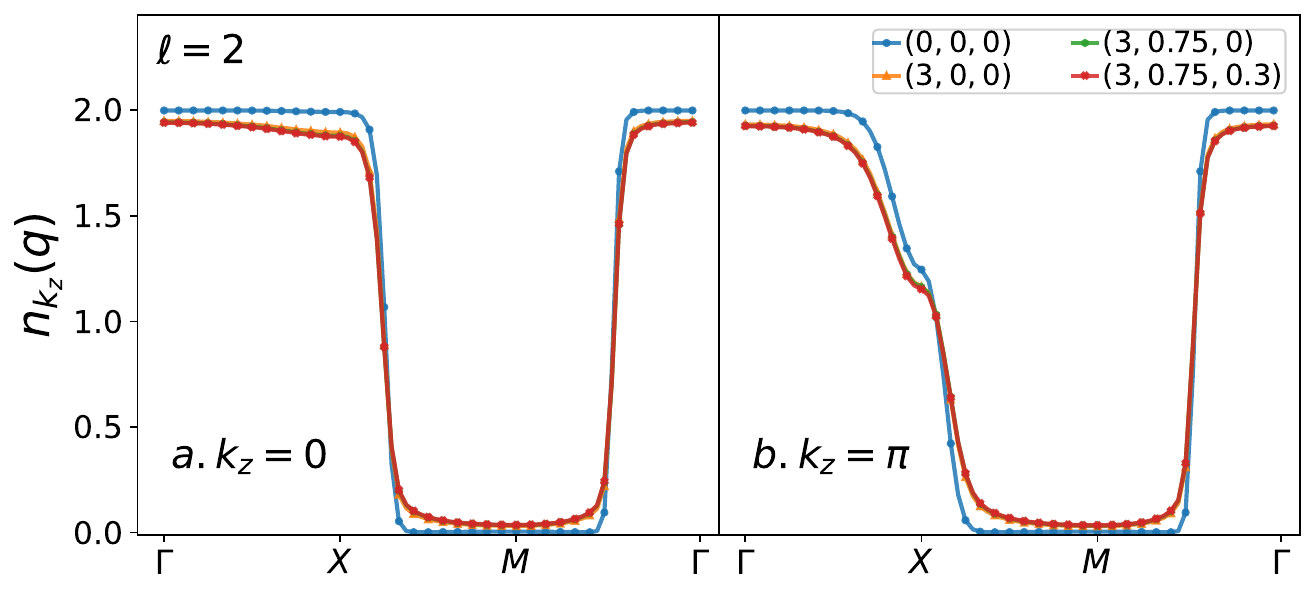}
  \caption{ Momentum distribution of spin summed filling  in  $\ell=2$ for  (a) $k_{z}=0$  and  (b) $k_{z} =\pi$ at $\mu=-0.6$ along the high symmetry cut in $\mathbf{q}$  where $\Gamma = [0,0]$, $X=[\pi,0]$ and $M=[\pi,\pi]$ for bare and interacting regimes.  Here, we set $J'/J =-0.5$  and $\beta=5$. }
  \label{fig:density_ell2}
\end{figure}

\begin{figure}[ht]
  \centering
  \includegraphics[width=1\linewidth]{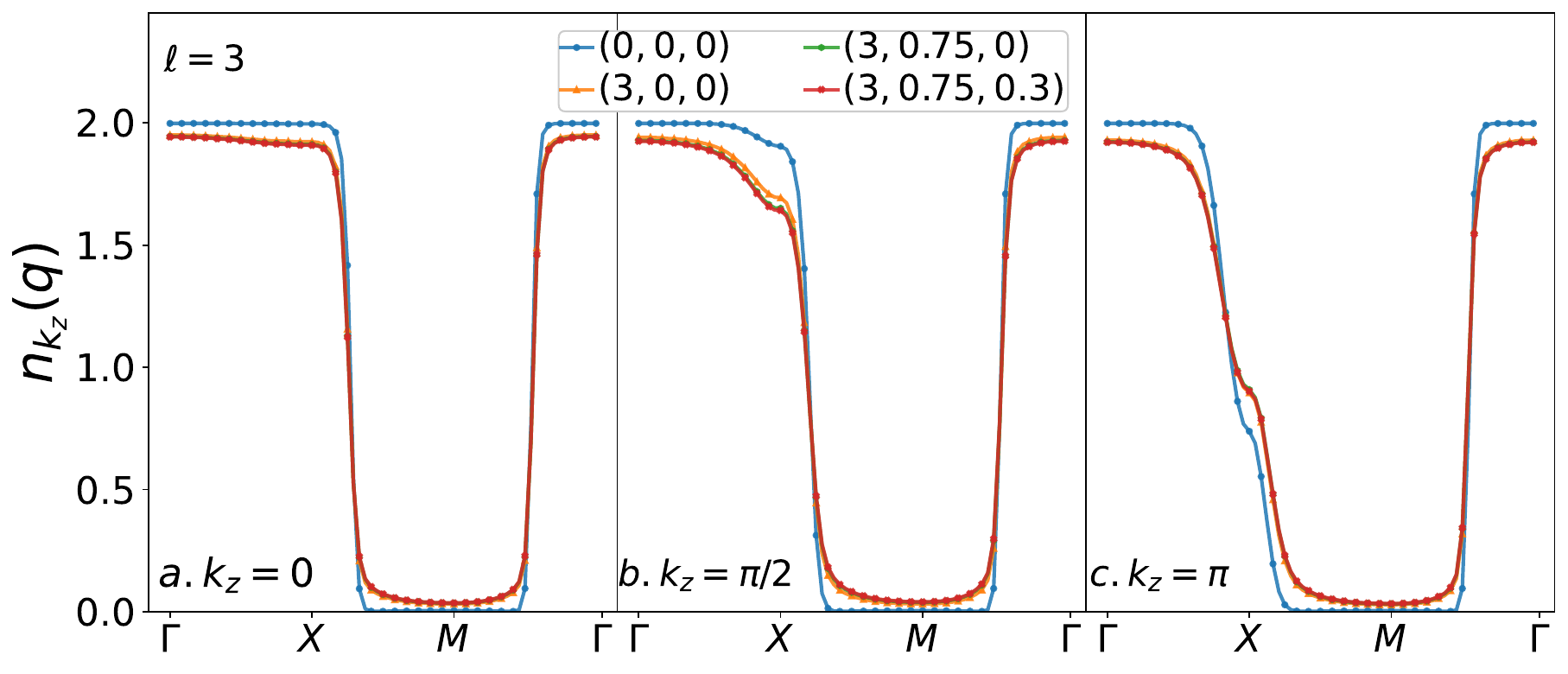}
  \caption{ Momentum distribution of spin summed filling in  $\ell=3$ at $\mu=-0.6$ for  (a)  outer plane $k_{z}=0$   (b) inner plane $k_{z} =\pi/2$  (c) outer plane $k_{z}=\pi$ along the high symmetry line in $\mathbf{q}$  for different choices of interaction parameter. Here, we set $J'/J =-0.5$  and $\beta=5$. }
  \label{fig:density_ell3}
\end{figure}

\begin{figure}[ht]
  \centering
  \includegraphics[width=1\linewidth]{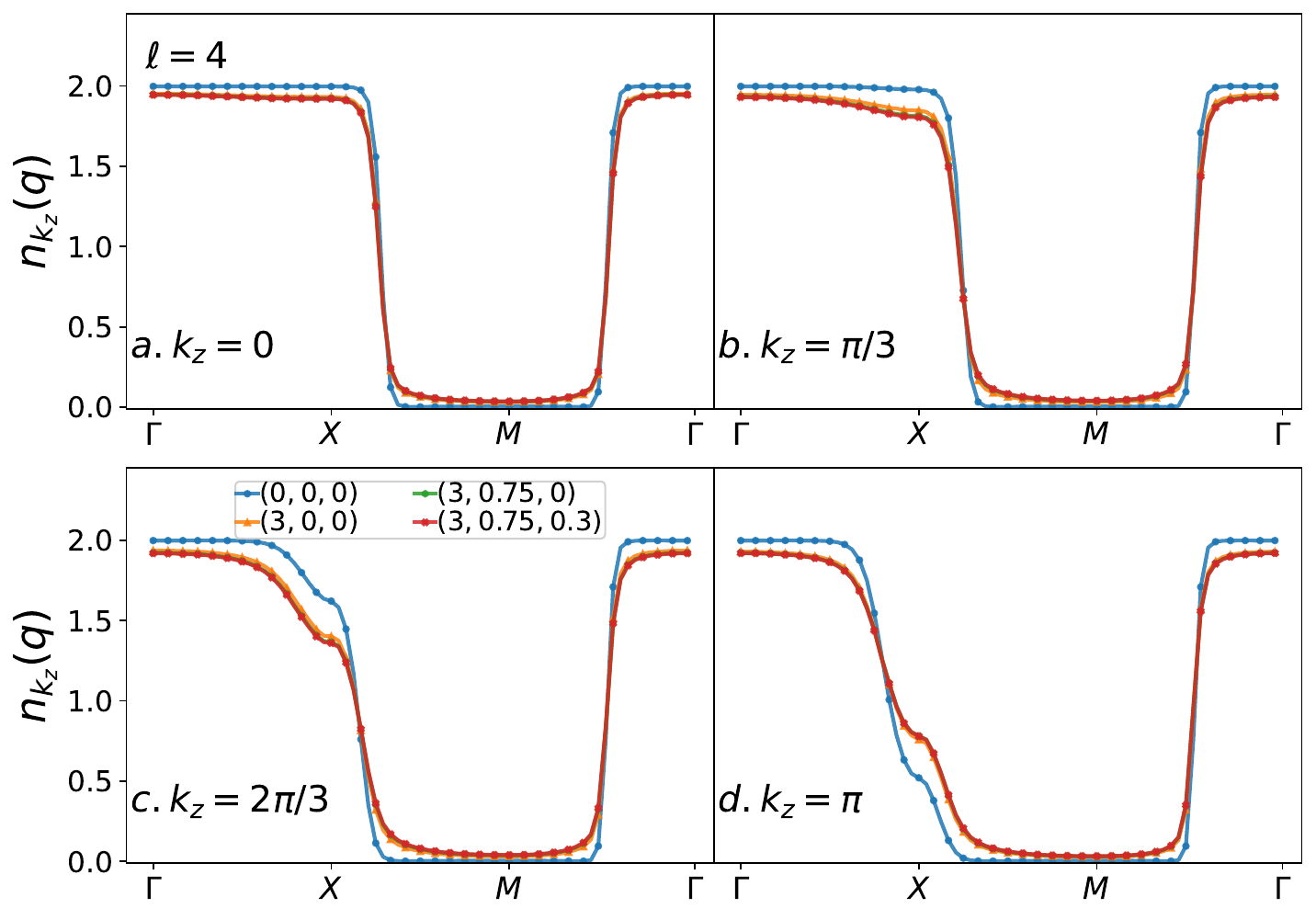}
  \caption{Momentum distribution of spin summed filling in  $\ell=3$ at $\mu=-0.6$ for  (a,d)  outer planes: $k_{z}=0$ and $k_{z}=\pi$  and  (b,c) inner planes: $k_{z} =\pi/3$ and  $k_{z} =2\pi/3$ along the high symmetry line in $\mathbf{q}$   for different choices of interaction parameter. Here, we set $J'/J =-0.5$  and $\beta=5$.  }
  \label{fig:density_ell4}
\end{figure}

\begin{table}[ht]
    \centering
    \renewcommand{\arraystretch}{1.1}
    \setlength{\tabcolsep}{10pt}
    \begin{tabular}{|c|c|c|c|}
    \hline
    \textbf{($U$, $V$, $J$)} & \textbf{$n_{0}$} & \textbf{$n_{\pi}$} & \textbf{$n_{\text{avg}}$}  \\
    \hline
    (0, 0, 0) & 1.10 & 0.920 & 1.01   \\
    \hline
    (3, 0, 0) & 1.06 & 0.915 & 0.99 \\ 
    \hline
    (3, 0.75, 0) & 1.06 & 0.915 & 0.99 \\ 
    \hline
    (3, 0.75, 0.3) & 1.06 & 0.915 & 0.99  \\
    \hline
    \end{tabular}
    \caption{\label{tab1} Spin summed  bare and renormalized density $n_{k_{z}}$ in $\ell=2$ system for same choice of parameters as the Fig.~\ref{fig:density_ell2}}
\end{table}
First, we inspect the spin-summed momentum distribution of the band filling $n_{k_{z}}(\mathbf{q})$ for $\ell=2,3$  and $4$ systems at $\mu=-0.6$. In the fermi liquid regime, the $n_{k_{z}}(\mathbf{q})$ is marked by a sharp discontinuity when approaching the fermi wavevector ($\mathbf{q} =k_{F}$) where the filling disappears. It also provides a measure for band renormalization and other strong correlation effects. We plot $n_{k_{z}}(\mathbf{q})$ along the high symmetry cut for $\ell=2,3$ and $4$ in Fig.~\ref{fig:density_ell2}(a-b), Fig.~\ref{fig:density_ell3}(a-c)  and  Fig.~\ref{fig:density_ell4}(a-d)  at $\beta=5$, respectively for a non-interacting case, and three choices of interaction parameters that is studied in main text. We have fixed the ratio, $J'/J=-0.5$ throughout.

\begin{table}[ht]
    \centering
    \renewcommand{\arraystretch}{1.1}
    \setlength{\tabcolsep}{10pt}
    \begin{tabular}{|c|c|c|c|c|}
    \hline
    \textbf{($U$, $V$, $J$)} & \textbf{$n_{0}$} & \textbf{$n_{\pi/2}$}  &  \textbf{$n_{\pi}$} & \textbf{$n_{\text{avg}}$}\\
    \hline
    (0, 0, 0) & 1.12 & 1.02 & 0.86 & 1.00 \\
    \hline
    (3, 0, 0) & 1.09 & 1.00 & 0.88 & 0.99 \\ 
    \hline
    (3, 0.75, 0) & 1.08 & 0.99  & 0.88 & 0.98 \\ 
    \hline
    (3, 0.75, 0.3) & 1.08 & 0.99 & 0.88 & 0.98  \\
    \hline
    \end{tabular}
    \caption{\label{tab2} Spin summed  density $n_{k_{z}}$ in $\ell=3$ system for same choice of parameters as the Fig.~\ref{fig:density_ell3}}
\end{table}
When a finite $U=3$ is introduced, we see a smearing of momentum distribution in $\ell=2,3$ and $4$ systems but the discontinuity still appears and remains sharp as in the bare regime when approaching near $k_{F}$. The lack of momentum redistribution can be pinned to the weakly momentum dependent nature of the $a_{[2,0,0]}$ coefficient in $\Sigma^{(2)}$. When $V$ and $J$ are introduced, no notable smearing is seen for low values of $V=0.75$ and $J=0.3$  utilized. However, the inner momenta planes appear to be more susceptible to V where  $k_{z} = 0$ for $\ell=3$  in Fig.~\ref{fig:density_ell3}(b) and for $k_{z}=\pi/3$ and $k_{z}=2 \pi/3$ in Fig.~\ref{fig:density_ell3}(b,c) exhibits small redistribution that is not seen in the outer layers.
\begin{table}[h!]
    \centering
    \setlength{\tabcolsep}{8pt}
    \begin{tabular}{|c|c|c|c|c|c|}
    \hline
    \textbf{($U$, $V$, $J$)} & \textbf{$n_{0}$} & \textbf{$n_{\pi/3}$}  &  \textbf{$n_{2\pi/3}$} &  \textbf{$n_{\pi}$}&\textbf{$n_{\text{avg}}$}\\
    \hline
    (0, 0, 0) & 1.13 & 1.07 & 0.97 & 0.84 & 1.000\\
    \hline
    (3, 0, 0) & 1.10 & 1.04 & 0.95  & 0.87 & 0.970\\ 
    \hline
    (3, 0.75, 0) & 1.09 & 1.03  & 0.94 & 0.87 & 0.965 \\ 
    \hline
    (3, 0.75, 0.3) & 1.09 & 1.03 & 0.94 & 0.87 & 0.965 \\
    \hline
    \end{tabular}
    \caption{ \label{tab3} Spin summed  density $n_{k_{z}}$ in $\ell=4$ system for same choice of parameters as the Fig.~\ref{fig:density_ell4}}
\end{table}
Summing the momentum resolutions, we obtain the total densities, which is summarized in Tab.~\ref{tab1}, Tab.~\ref{tab2}, and Tab.~\ref{tab3} for $\ell=2,3$ and 4,  systems respectively. We see a trend where the inclusion of $U$ weakens doping while $V$ only affects the inner momenta layers. Both $J$ and $J'$ yield no discernible changes to doping. The renormalized doping in the second order remains comparable to the bare densities within the weak coupling regime.

\subsection{Dependence of $t_{z}$ and $t_{bs}$  in bilayer system}

\begin{figure}[ht]
    \centering
    \includegraphics[width=1\linewidth]{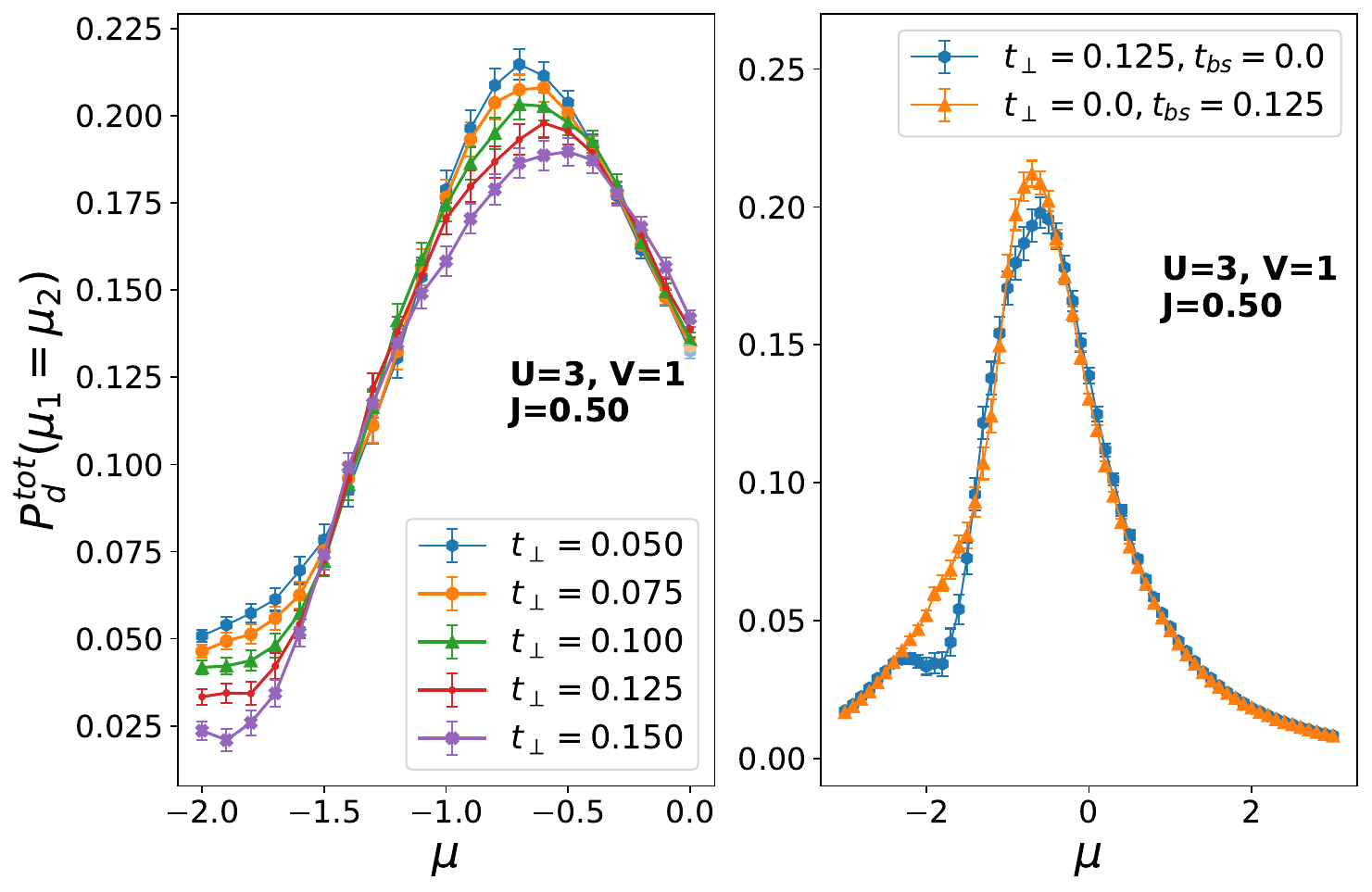}
    \caption{\label{iso_aniso} (left panel) Anisotropic tunneling strength dependence (right panel) comparison between isotropic and anisotropic tunneling strength of intra-layer $d_{x^2-y^2}$ pairing susceptibility in Bilayer Hubbard model for equal $\mu_{1}=\mu_{2}$ case at $\beta=5t$ and $J'/J=-0.5$. }
\end{figure}

Here we analyze the role  of anisotropic $t_{\perp}(k)$ tunneling strength on $P^{tot}_{d}$ and compare it with the isotropic $t_{bs}$ with $t_{\perp}$. To do so, we narrow down the parameter space of $\vec{\mu}$ by setting the chemical potentials of the $k_{z} = 0$ band ($\mu_{1}$) and $k_{z}=\pi$ band ($\mu_{2}$) equal to each other $\mu_{1} = \mu_{2}$. In the Fig.~\ref{iso_aniso}(a), we present $P^{tot}_{d}$ as a function of chemical potential $\mu$ with five choices of $t_{\perp}$ ranging from $0.05$ to $0.15$ with incremental steps of $0.025$ for $U=3$, $V=1$, and $J=0.50$. We find that the peak resides on the $\mu <0$ side, and increasing $t_{\perp}$ has an overall adverse effect on $P^{tot}_{d}$. In Fig.~\ref{iso_aniso}(b), we contrast the influence of isotropic and anisotropic splitting as a function of $\mu$. By noting that $t_{\perp}(k)$ produces an average  splitting  equal to magnitude $t_{\perp}$, (since $\frac{1}{4\pi^2} \int_{0}^{2\pi}\int_{0}^{2\pi} \left[\cos(x) - \cos(y)\right]^2 \, dxdy = 1$ ), one can directly compare with results of $t_{bs}$. We find that for the choice of parameters  presented,  $(t_{bs},t_{\perp})=(0.0,0.125), (0.125,0.0)$, the picture remains relatively unchanged with $t_{bs}$ case seeing a small enhancement. Nevertheless, as the strength of both $t_{bs}$ and $t_{\perp}$ increases, pairing is weakened monotonically and eventually, for sufficiently large enough value $P^{tot}_{d}$ is fully quenched. This pattern remains consistent when probed across $U,V,J$ interaction space up to $\ell =4$ system we study (not shown). This relation has been well documented in several bilayer studies with isotropic $t_{bs}$ tunneling where suppression of $d_{x^2-y^2}$-wave channel is usually accompanied by the emergence of $d_{z}$-wave channel\cite{isotropic(scalepino),iso(Karakuzu)}.

\subsection{$P^{tot}_{d}$ in the unconstrained $\mu$ phase space for bilayer and trilayer system}

\begin{figure}[ht]
    \centering
    \includegraphics[width=1\linewidth]{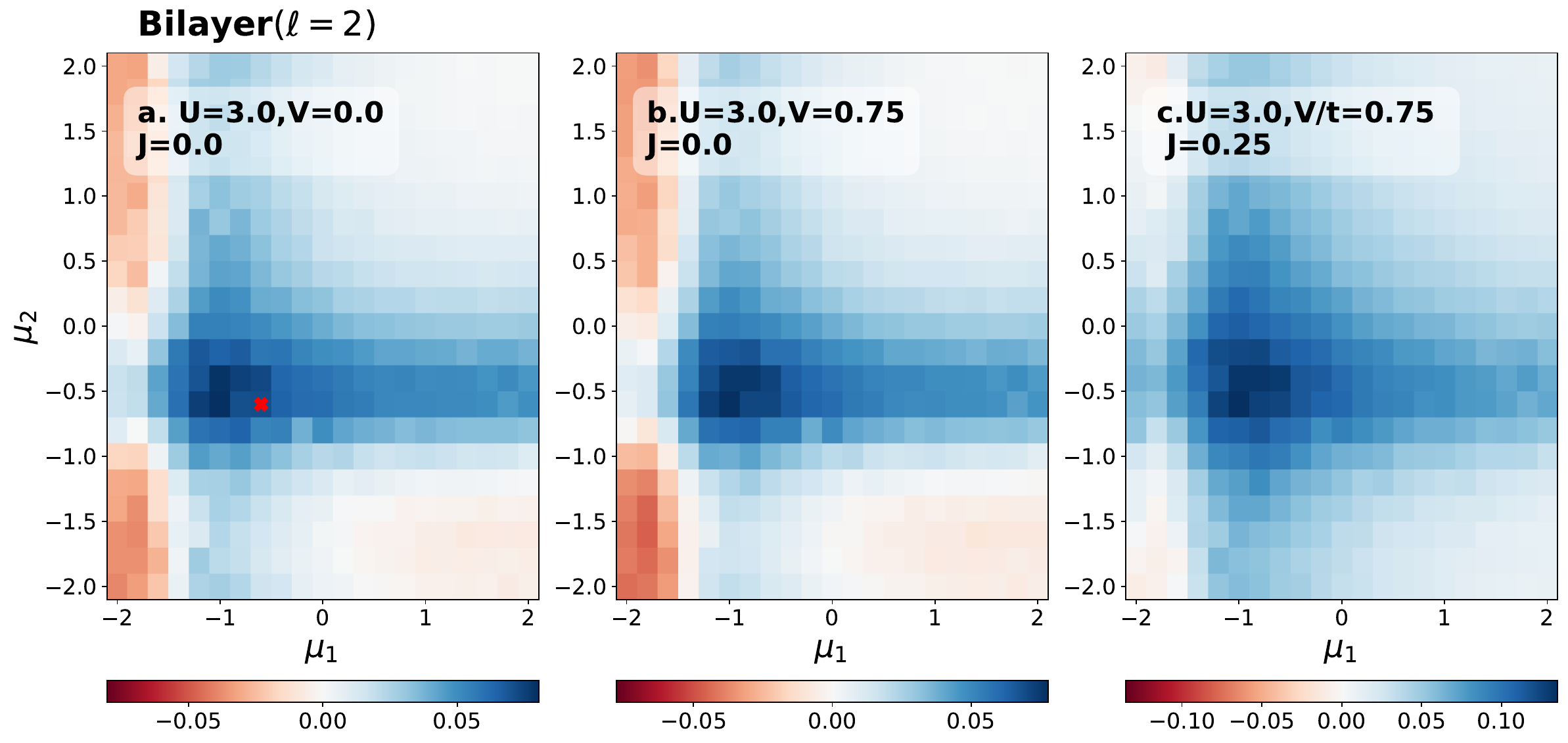}
     \includegraphics[width=1\linewidth]{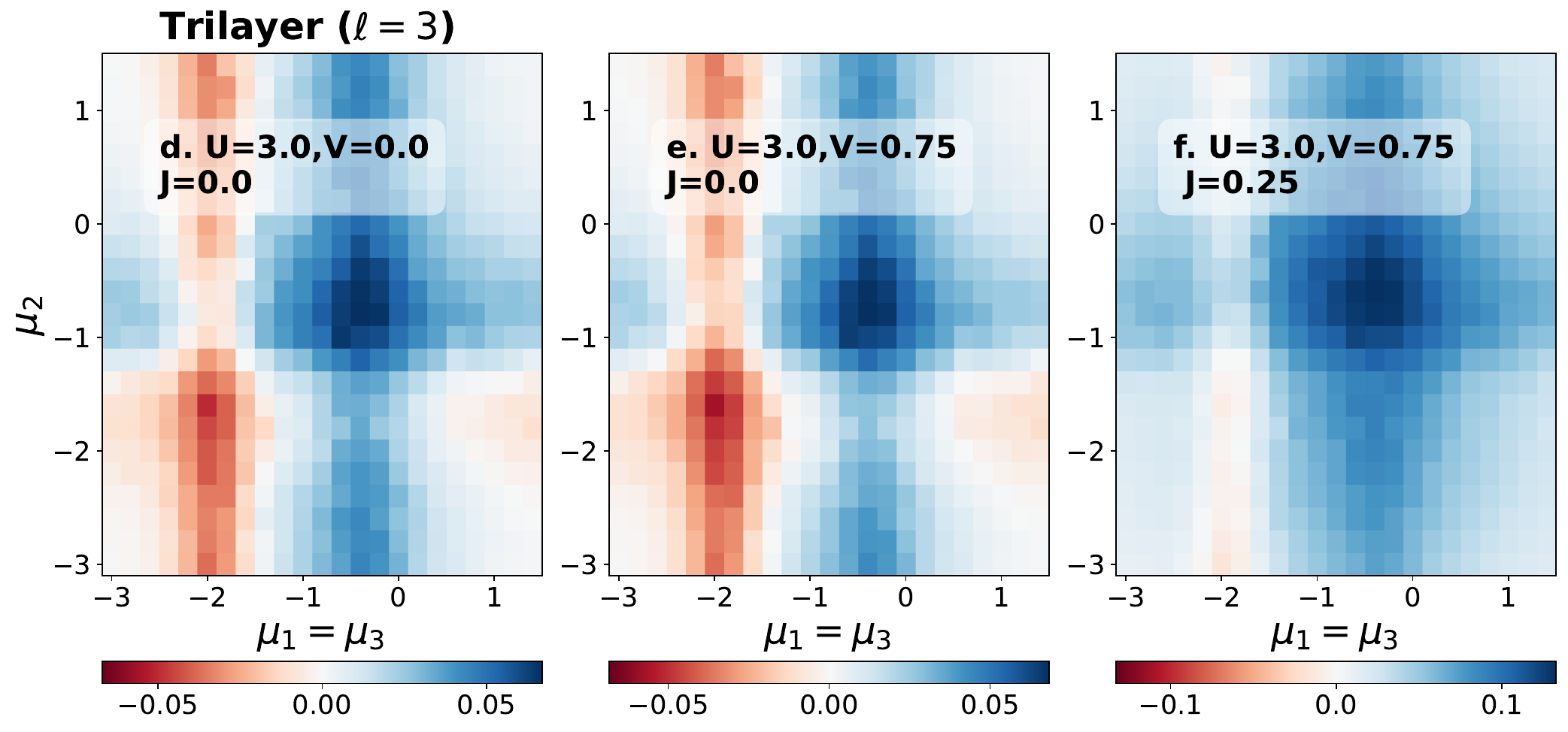}
    \caption{\label{mu_mu_bilayer} False color plots of $P^{tot}_d$ in bilayer (a-c) and trilayer (d-f) models for three choices of interaction sets. The bilayer model treats $\mu$ of bonding $\mu_{1}$ and antibonding ($\mu_{2}$) as independent variables. Red cross indicates the peak along $\mu_{1}=\mu_{2}$ line. In the trilayer model, the $\mu$ of the outer momenta planes $k_{z}=0$ $(\mu_{1})$ and $k_{z}=\pi$ $(\mu_{3})$ are set equal and varied  with inner momenta plane  $k_{z}=0$ $(\mu_{2})$. We fix $t_{\perp}=0.125, t_{bs}=0$, $J'/J=-0.5$, $\beta=5t$ in each case.}
\end{figure}

In this section, we lift the previously imposed constraint of the $\mu_{1} = \mu_{2}$ and present false-color plots of $P^{tot}_{d}$ in Fig.~\ref{mu_mu_bilayer}(a-c) at $U=3$ for three choices of  $V$ and $J$.  It is well established that van Hove singularity (VHS), corresponding to a saddle point in single particle dispersion, plays a key role in $d_{x^2-y^2}$-wave pairing \cite{VHS}. We have demonstrated in our prior single band study that the pairing is largest when it is in proximity to VHS but experiences a heavy suppression exactly at and beyond VHS occurring at  $\mu=4t^{\prime}$\cite{farid:2023}.  By allowing the $\mu$ in the bands to be independently controlled, one can fine-tune the fermi surfaces such that they are close to VHS, thereby enhancing pairing.

First, we focus on the bilayer system. In Fig.~\ref{mu_mu_bilayer}(a) with $V, J=0$, the $P^{\perp}_{d}$ term is absent and two diagonal channels $P^{0,0}_{d}$ and $P^{\pi,\pi}_{d}$  can be evaluated as separate single-band cases. Since the propagators in truncated third-order $P^{tot}_{d}$ vertex diagrams can only be normalized by a weakly momentum dependent second-order self-energy at most, we expect pairing to be dictated by the location of non-interacting VHS. We present the location of van Hove singularity in $\ell=2$ system in Fig.~\ref{fig:fermi}(a). We find that the optimal $\mu$ occurs at $(\mu_{1},\mu_{2})=-(1.2,0.6)$  closely following the proximity to VHS located at $-(1.7,0.7)$ but the $P^{tot}_{d}$ strength is only marginally enhanced approximately by  $15\%$ in comparison to the peak along $\mu_{1}=\mu_{2}$ line as indicated by the red cross at $\mu=-0.6$. With the introduction of finite $V$ and $J$ values in Fig.~\ref{mu_mu_bilayer}(b,c), several particle-particle diagrams corresponding will have propagators with different band indices with distinct VHS, in contrast to the $V,J=0$ case. One might expect this to shift in the optimal $\mu$ region. However, the qualitative features remain mostly the same apart from changes in $P^{tot}_{d}$ magnitude.  It's worth noting that a wide region exists where two bands have chemical potentials with opposite signs with attractive $P^{tot}_{d}$, albeit weaker than the optimal region. The $P^{tot}_{d}(\mu_{1}=-\mu_{2})$   case, in particular, has been studied through a Quantum Monte Carlo simulation in a bilayer model with isotropic tunneling $t_{bs}$ in relevance to four-layer high $T_{c}$ Ba$_2$Ca$_3$Cu$_4$O$_8$F$_2$ cuprate where the two outer and two inner planes in real space are oppositely electron and hole-doped \cite{iso(Bouadim),exp(hole_electro)}.

As investigated in detail in the main text, the presence of a small $J=0.25$ that is comparable to Hund's coupling leads to an overall enhancement of $P^{tot}_{d}$ over the entire $\mu_{1}-\mu_{2}$  phase. Around the peak, an enhancement by an approximate factor of two is observed when compared to $J=0$ case. Further, we see that the inclusion of $J$ results in a more prominent enhancement in Fig.~\ref{mu_mu_bilayer}(c) compared to just approaching the optimal $\vec{\mu}$ region for  $J=0$ case in Fig.~\ref{mu_mu_bilayer}(a).  

 We now focus on probing  $\vec{\mu}$ space in the trilayer system. Reiterating our earlier remark, the optimal $\vec{\mu}=(\mu_{1},\mu_{2}$,$\mu_{3})$ space for the $k_{z}=0,\pi/2,\pi$  bands are also expected to follow the VHS shown in Fig.~\ref{fig:fermi}. However, we encounter a large computational hurdle associated with probing the full three-dimensional $\vec{\mu}$ space. To simplify this, we set the chemical potential in the bonding and anti-bonding planes equal $\mu_{1}=\mu_{3}$ and along with $\mu_{2}$ probe the $\mu_{1,3}-\mu_{2}$ space. In Fig.~\ref{mu_mu_bilayer}(d-f), we present the false color plots for $P^{tot}_{d}$ with three choices of interaction and identify the optimal $(\mu_{2}, \mu_{1,3})$ where $P^{tot}_{d}$ is largest. We find that the overall picture remains largely unchanged from the bilayer case, where the role $J$ is more prominent than approaching optimal $\vec{\mu}$ region. However, when comparing the pairing between $\ell=2$ and $\ell=3$ at $J=0$, we observe that the pairing in the optimal region for $\ell=3$  is slightly attenuated compared to $\ell=2$. The three panels in Fig.~\ref{mu_mu_bilayer}(d-f)  features a region of attractive pairing in $\ell=3$ system peaked at $(\mu_{2}, \mu_{1,3})= -(0.8,0.4)$. We see that optimal $\mu$ also follows the location of VHS in $k_{z}=\pi/2$ and $k_{z} = \pi$ bands located  at $\mu_{2}=-1.2$ and $\mu_{3}=-0.5$. The pairing towards the  VHS in $k_{z}=0$ band, located at $\mu_{1}=-1.9$, is not preferred as this would require going beyond VHS of both $k_{z}=\pi/2, \pi $  bands resulting in  $P^{\pi/2,\pi/2}_{d}$ and $P^{\pi,\pi}_{d}$ channels becoming repulsive, thereby suppressing overall pairing. This is evident from the pronounced repulsive region beyond  $\mu_{1,3}<-1.6$. Our results indicate that it's possible for both $\ell=2,3$ systems to take advantage of the VHS positions and $J$ interaction to enhance $P^{tot}_{d}$. We provide the corresponding  $n_{k}$ density in the optimal $\mu$ region from Fig.~\ref{mu_mu_bilayer} in Tab.~\ref{tab4} for $\ell=2$ 
 and in Tab.~\ref{tab5} for $\ell=3$.
\begin{table}[ht]
    \centering
    \renewcommand{\arraystretch}{1}
    \setlength{\tabcolsep}{10pt}
    \begin{tabular}{|c|c|c|c|}
    \hline
    \textbf{($U$, $V$, $J$)} & \textbf{$n_{0}$} & \textbf{$n_{\pi}$} & \textbf{$n_{\text{avg}}$} \\
    \hline
    (0, 0, 0) & 0.875 & 0.920 & 0.90  \\
    \hline
    (3, 0, 0) & 0.860 & 0.915 & 0.89  \\ 
    \hline
    (3, 0.75, 0) & 0.860 & 0.915 & 0.89   \\ 
    \hline
    (3, 0.75, 0.25) & 0.860 & 0.915 & 0.89  \\
    \hline
    \end{tabular}
    \caption{\label{tab4} Spin summed density corresponding to the optimal $(\mu_1,\mu_2)=-(1.2,0.6)$ region in  $\ell=2$ from Fig.~\ref{mu_mu_bilayer}(a-c) a. Here $\mu_{1}$ and $\mu_{2}$ corresponds to chemical potential in $k_{z}=0$ and $k_{z}=\pi$ momenta layers, respectively. }
\end{table}
\begin{table}[ht]
    \centering
    \renewcommand{\arraystretch}{1}
    \setlength{\tabcolsep}{10pt}
    \begin{tabular}{|c|c|c|c|c|}
    \hline
    \textbf{($U$, $V$, $J$)} & \textbf{$n_{0}$} & \textbf{$n_{\pi/2}$}  &  \textbf{$n_{\pi}$} & \textbf{$n_{\text{avg}}$} \\
    \hline
    (0, 0, 0) & 1.175 & 0.950 & 0.96 & 1.03\\
    \hline
    (3, 0, 0) & 1.140 & 0.925 & 0.95 & 1.00 \\ 
    \hline
    (3, 0.75, 0) & 1.135 & 0.920  & 0.95 & 1.00 \\ 
    \hline
    (3, 0.75, 0.3) & 1.135 & 0.920 & 0.95 & 1.00 \\
    \hline
    \end{tabular}
    \caption{\label{tab5}  Spin summed density corresponding to the optimal $(\mu_{2},\mu_{1,3}) =-(0.8,0.4)$ region in  $\ell=3$ from Fig.~\ref{mu_mu_bilayer}(d-f). Here, $\mu_1,\mu_2,$  and $\mu_3$ corresponds to chemical potential in $k_{z}=0$, $k_{z}=\pi/2$, and  $k_{z}=\pi$ momenta layers respectively. } 
\end{table}

\subsection{Coefficients $a_{[i,j,k]}$ in trilayer system}
\begin{figure}[ht]
    \centering
    \includegraphics[width=1\linewidth]{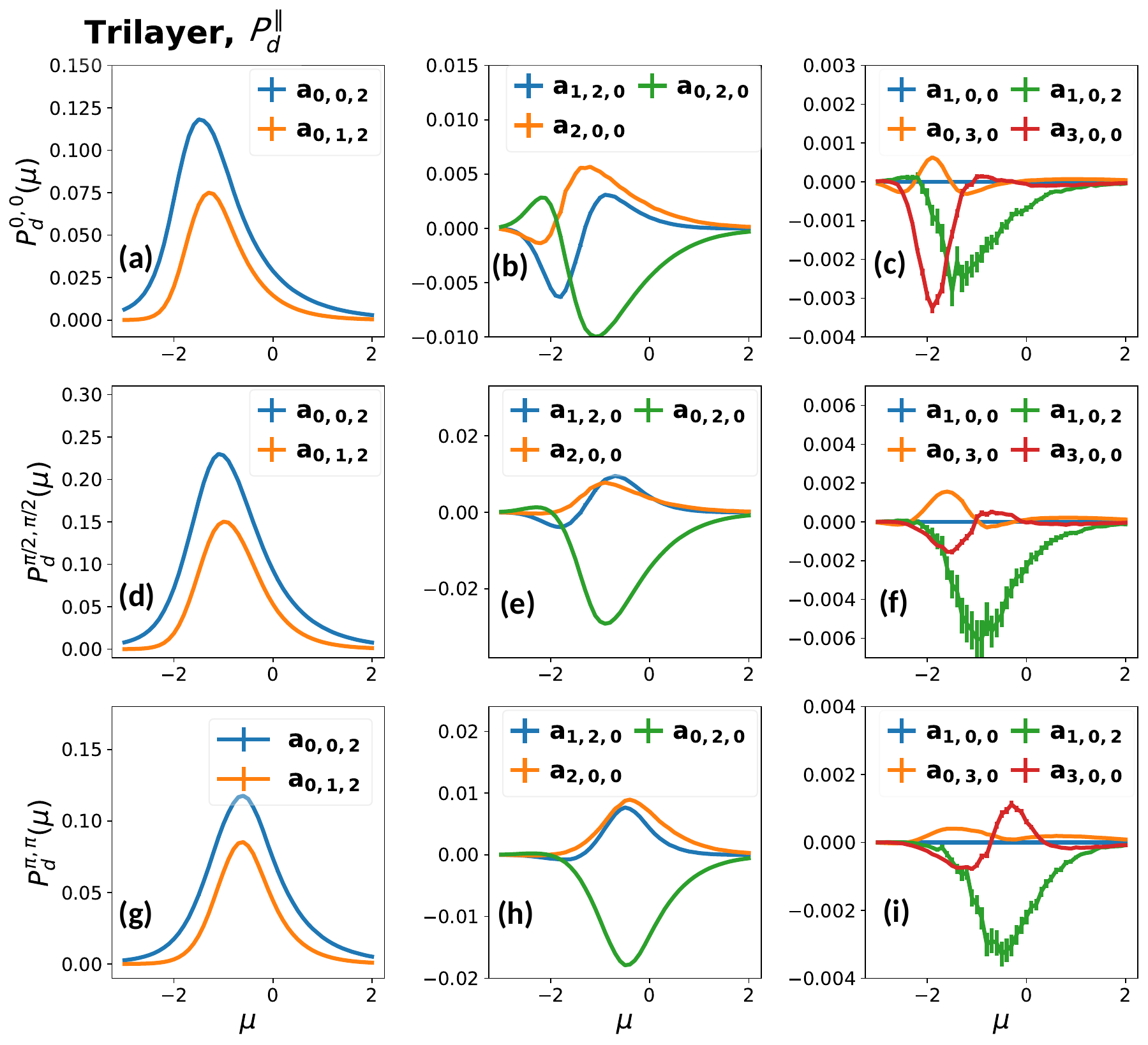}
    \caption{ \label{coeff_diag}The coefficients of the three inequivalent diagonal channels are provided as a function of $\mu$ in $\ell=3$  system for $U,V,J=1$ with $J'/J=-0.5$.}
 \end{figure}
\begin{figure}[ht]
    \centering
    \includegraphics[width=1\linewidth]{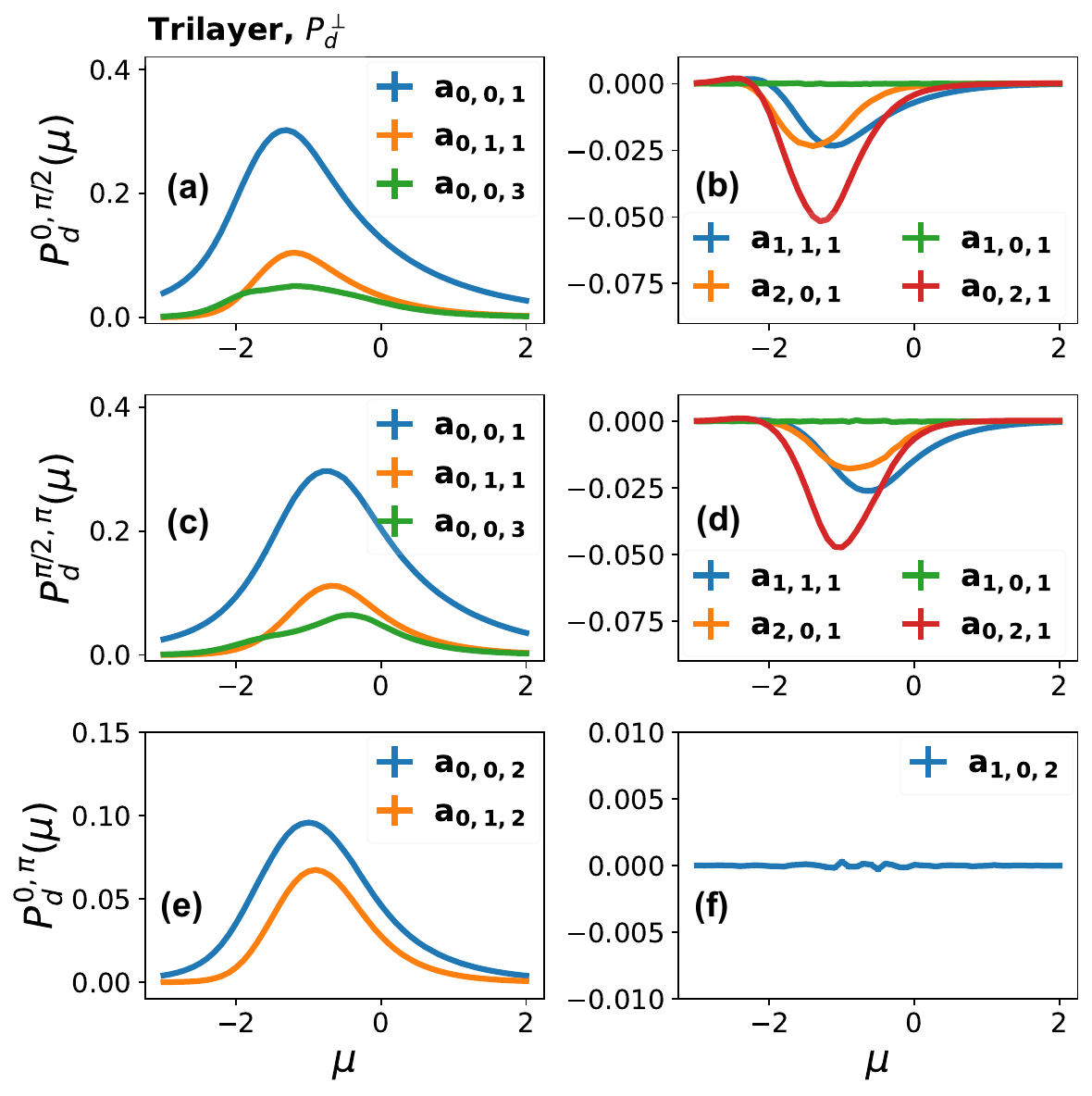}
    \caption{\label{coeff_ofdiag}The coefficients of the three inequivalent off-diagonal channels are provided as a function of $\mu$ in $\ell=3$  system for $U,V,J=1$ with $J'/J=-0.5$.}
 \end{figure}

We provide the truncated third order coefficients for $\ell=3$ system for the three inequivalent diagonal channels in Fig.~\ref{coeff_diag} and off-diagonal channels in Fig.~\ref{coeff_ofdiag} as a function of $\mu$ for $U,V,J=1$ with a fixed ratio of $J'/J=-0.5$. Here, the coefficients are categorized in the order of magnitude, with the first column being the dominant contributor and the last column being the least. The dominant coefficients in the diagonal channels are shared with those in the diagonal channels of the $\ell=2$ system. The magnitude of the coefficients is relatively similar, but the location of the peak varies in $\mu$ space, owing to the difference in VHS position in each band. Among the three channels, we observe that $P^{\pi,\pi}_{d}$ has the largest contributor to $P^{\parallel}_{d}$. This is due to comparatively larger contribution from $a_{[2,0,0]}$ and $a_{[3,0,0]}$ coefficients, which we have shown in the main text to be the dominant contributor at $U=3$.

Finally, two adjacent inequivalent off-diagonal channels in Fig.~\ref{coeff_ofdiag}(a-d) share the same dominant coefficient as the $\ell=2$ system. The only difference is the presence of the next adjacent off-diagonal channel given by $P^{0,\pi}_{d}$, which has two positive non-zero  $a_{[0,0,2]}$ and $a_{[0,1,2]}$ as shown in Fig.~\ref{coeff_ofdiag}(e,f).

\subsection{Spin correlation in bilayer system}

From the linear response theory, intra-layer, the static  $\Omega=0$ spin  susceptibility in the bilayer Hubbard model  can be calculated via
\begin{equation}
\chi^{k_{z},k'_{z}}_{s}(\mathbf{q})= \int^{\beta}_{0} d\tau \langle S^{k_{z}}_{z} (\mathbf{q},\tau) S^{k'_{z}}_{z}(\mathbf{-q},0)\rangle 
\vspace{0.1cm}
\end{equation}

where $S^{k_{z}}_{z} (\mathbf{q},\tau)$ and $S^{k'_{z}}_{z} (\mathbf{-q},0)$ are the spin operators in the $k_z$ and $k'_z$ orbital channels, respectively. Note that $\chi^{k_{z},k'_{z}}_{s}$ with $k_{z} \neq k'_{z}$  requires a finite V interaction to have any contribution.
 
 The total susceptibility can be calculated by simply adding all the spin channels normalized by $\ell$, resulting in
\begin{equation}
    \chi^{tot}_{s} = \frac{1}{2}\sum_{k_{z},k'_{z}}\chi^{k_{z},k'_{z}}_{s}
\end{equation}
\begin{figure}[ht]
    \centering
    \includegraphics[width=1\linewidth]{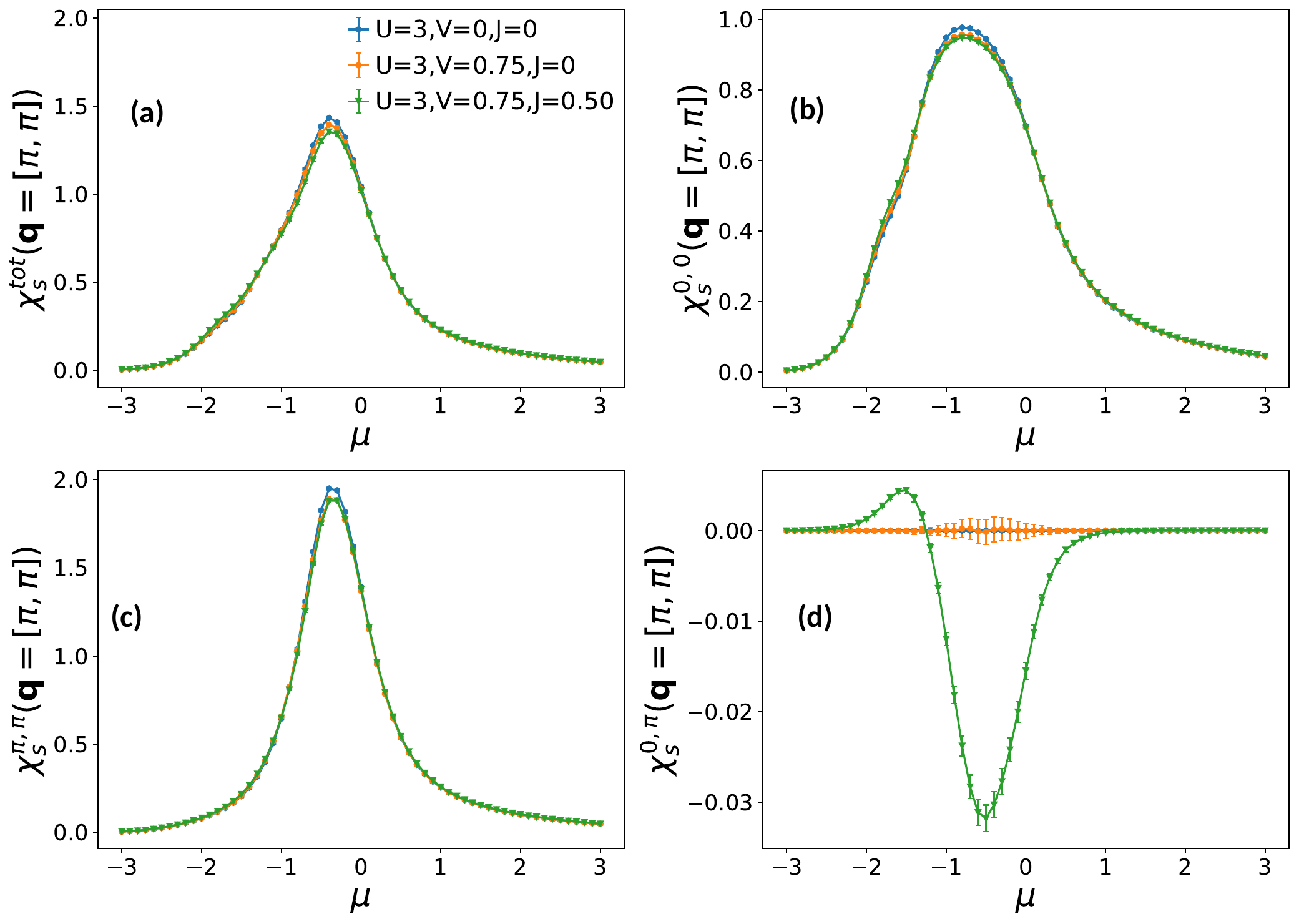}
    \caption{\label{spin} Total spin susceptibility and spin channel as a function of $\mu$ at $\beta =5$ and $J'/J=-0.5$ for three choices of $U$,$V$ and $J$ interactions. }
 \end{figure}

The diagrammatic expansion of $\chi^{k_{z},k'_{z}}_{s}$  can be similarly performed via symbolic multiband sampling as the $P^{k_{z},k'_{z}}_{g}$ case, as outlined in the main text, where we obtain a set of $a_{[i,j,k]}$ coefficient and perform multi-series expansion. Here, we primarily study the $(q_{x},q_{y})=(\pi,\pi)$ case, where antiferromagnetic spin fluctuation   $\chi^{k_{z},k'_{z}}_{s}(\mathbf{q})$ is peaked and is known to be the key mediator of $d$-wave anomalous self-energy.\cite{Dong}

In order to demonstrate the role of $U,V,J$ interaction, we plot truncated third-order total spin susceptibility $\chi^{total}_ {s}$ in Fig.~\ref{spin}(a) and its three inequivalent spin channels $\chi^{k_{z},k'_{z}}_{s}$ in  Fig.~\ref{spin}(b-d) as a function of $\mu$ for three choices of interactions that is used extensively in the main text. Firstly, we immediately see that diagonal $\chi^{\pi,\pi}_{s}$ has the largest contribution, which is twice as large as the next leading term $\chi^{0,0}_{s}$. This is in agreement with the trends pairing where $P^{\pi,\pi}_{d} > P^{0,0}_{d}$ in $\ell=2$. The off-diagonal $\chi_{s}^{0,\pi}$ features zero contribution at $V=0$ and surprisingly with finite $V=0.75$ as well. The latter can be attributed to the cancellation of two density-density terms $\chi_{\uparrow\uparrow}$ and $\chi_{\uparrow\downarrow}$ in the off-diagonal channel where $\chi_{s}=\chi_{\uparrow\uparrow}-\chi_{\uparrow\downarrow} $. With finite $J=0.5$, there is a finite negative contribution in $\chi_{s}^{0,\pi}$, albeit significantly smaller when compared to the diagonal channels. In the diagonal channel, the inclusion of $V$ results in weak suppression, while the inclusion of $J$ has no effect. As a result, there is only a marginal suppression in $\chi^{tot}_{s}$ via the inclusion of  $V$ and  $J$  due to the weakening of the diagonal and off-diagonal channels, respectively. This is in contrast to $P^{tot}_{d}$, where we have shown in the main text that there is a quadratic increase with $J$ when $J'/J<0$.
\bibliography{refs.bib}

\end{document}